\documentclass[prd,preprint,tightenlines,11pt,amsmath,amssymb]{revtex4}

\usepackage{graphicx}       
\usepackage{dcolumn}        
\usepackage{multirow}
\usepackage{bm}             
\usepackage{psfrag}
\usepackage{color}

\newcommand{\beq}{\begin{equation}}
\newcommand{\eeq}{\end{equation}}

\newcommand{\Phires}{\Phi_{\mathcal{R}}}

\newcommand{\Psires}{\Psi_{\mathcal{R}}}

\newcommand{\Psireshat}{\hat{\Psi}_{\mathcal{R}}}

\newcommand{\Phipunc}{\Phi_{\mathcal{P}}}

\newcommand{\Psirestil}{\hat{\Psi}_{\mathcal{R}}}
\newcommand{\Psirettil}{\hat{\Psi}}

\newcommand{\Seff}{S_{\text{eff}}}

\newcommand{\dth}{\delta \theta}

\newcommand{\Phim}{\Phi^m}
\newcommand{\Psim}{\Psi^m}
\newcommand{\tmax}{t_\text{max}}

\newcommand{\mmax}{m_{\text{max}}}

\mathchardef\mhyphen="2D

\providecommand{\e}[1]{\ensuremath{\times 10^{-#1}}}

\newcommand{\EE}{\mathcal{E}}
\newcommand{\LL}{\mathcal{L}_z}

\newcommand{\bu}{\mathbf{u}}
\newcommand{\bF}{\mathbf{F}}
\newcommand{\bk}{\mathbf{k}}

\newcommand{\alp}{\alpha}

\newcommand{\gam}{\gamma}

\newcommand{\Delt}{\triangle t}
\newcommand{\Delr}{\triangle r_\ast}
\newcommand{\Delq}{\triangle \theta}

\providecommand{\e}[1]{\ensuremath{\times 10^{-#1}}}
\newcommand{\msp}{\phantom{-}}
\newcommand{\risco}{r_{\text{isco}}}

\newcommand{\sco}{\mathsf{s}}
\newcommand{\esq}{\mathsf{s}}
\newcommand{\rhohat}{\hat{\rho}}

\begin{document}

\preprint{}

 \title{Self force via $m$-mode regularization and 2+1D evolution: II. \\ 
 Scalar-field implementation on Kerr spacetime}

\author{Sam R. Dolan}
 \email{s.dolan@soton.ac.uk}
 \affiliation{%
 School of Mathematics, University of Southampton, Southampton SO17 1BJ, United Kingdom. \\
 }
\author{Barry Wardell}
 \email{barry.wardell@aei.mpg.de}
 \affiliation{
  Max-Planck-Institut f\"ur Gravitationsphysik, Albert-Einstein-Institut, Am M\"uhlenberg 1, D-14476 Golm, Germany.
 }
\author{Leor Barack}
 \email{l.barack@soton.ac.uk}
 \affiliation{%
 School of Mathematics, University of Southampton, Southampton SO17 1BJ, United Kingdom. \\
 }%

\date{\today} 

\begin{abstract}
This is the second in a series of papers aimed at developing a practical time-domain method for self-force calculations in Kerr spacetime. The key elements of the method are (i) removal of a singular part of the
perturbation field with a suitable analytic ``puncture'' based on the Detweiler--Whiting decomposition,
(ii) decomposition of the perturbation equations in azimuthal ($m$-)modes, taking advantage of the axial symmetry of the Kerr background, (iii) numerical evolution of the individual $m$-modes in 2+1-dimensions with a finite difference scheme, and (iv) reconstruction of the physical self-force from the mode sum. Here we report an implementation of the method to compute the scalar-field self-force along circular equatorial geodesic orbits around a Kerr black hole. This constitutes a first time-domain computation of the self force in Kerr geometry. Our time-domain code reproduces the results of a recent frequency-domain calculation by Warburton and Barack, but has the added advantage of being readily adaptable to include the back-reaction from the self force in a self-consistent manner. In a forthcoming paper---the third in the series---we apply our method to the gravitational self-force (in the Lorenz gauge). 
\end{abstract}

\pacs{}
\maketitle

%
%
%
%

\section{Introduction and overview\label{sec:intro}}

The context of this work is the ongoing program to apply the theory of self forces (SFs) in curved spacetime to the problem of a massive compact body orbiting a black hole \cite{Poisson:Pound:Vega:2011,Barack:2009}. This program is now reaching fruition. Recent highlights include a calculation of the gravitational SF along bound (but otherwise generic) geodesic orbits in Schwarzschild geometry \cite{Barack:Sago:2010}, the quantitative determination of several gauge-invariant post-geodesic effects associated with the gravitational SF \cite{Detweiler:2008,Barack:Sago:2009,Barack:Sago:2011}, and the first meaningful comparisons with post-Newtonian results \cite{Blanchet:Detweiler:LeTiec:Whiting:2010a,Blanchet:Detweiler:LeTiec:Whiting:2010b,Barack:Damour:Sago:2010}. Results from SF calculations are being used to inform the development of an Effective One Body (EOB) model of the inspiral \cite{Damour:2009,Barack:Damour:Sago:2010}, and even as a benchmark for fully-nonlinear numerical-relativistic simulations \cite{NR}. Despite this significant progress, gravitational SF study so far has been restricted to the relatively simple model where the background geometry is that of a non-rotating (Schwarzschild) black hole. In order for the program to become truly astrophysically relevant \cite{Barack:Cutler:2004}, the community must now come to address the problem of the gravitational SF in Kerr geometry.

A first calculation of the SF in Kerr spacetime has in fact been presented recently \cite{Warburton:Barack:2010,Warburton:Barack:2011}, employing the scalar-field SF as a (relatively) simple toy model for the gravitational SF itself. This calculation---first performed for a circular orbit in the equatorial plane of the Kerr black hole \cite{Warburton:Barack:2010} and later extended to eccentric orbits (still in the equatorial plane) \cite{Warburton:Barack:2011}---is based on a frequency-domain treatment of the field equations: the (scalar field) perturbation equations are integrated numerically mode-by-mode in a Fourier-harmonic decomposition, and the physical scalar-field SF is then constructed using standard mode-sum regularization \cite{Barack:Ori:2000}. This approach is computationally efficient and relatively easy to implement (as it reduces the computational task to, essentially, the solution of ordinary differential equations). However, the method also has several drawbacks: it becomes rapidly less efficient with increasing eccentricity; it cannot be incorporated easily in a scheme that evolves the orbit self-consistently under the back-reaction effect of the SF; and it is not easily generalizable to the gravitational SF in Kerr. (An alternative frequency-domain approach, designed specifically to deal with the last of the above challenges, is being developed by Shah {\it et al}.\ \cite{Keidl:Shah:Friedman:2010, Shah:Keidl:Friedman:2011}.)

Time-domain methods offer a natural way around the above difficulties. Numerical codes based on time evolution are versatile and flexible, and can accommodate different orbital configurations with relatively little alteration. They can, in principle, be readily adapted to incorporate the back reaction from the SF on the orbit in a self-consistent manner. The metric perturbation equations in the Lorenz-gauge lend themselves most naturally to a time-domain treatment thanks to their hyperbolic nature. Lastly, time-domain codes formulated in 2+1D or 3+1D do not rely on separability of the perturbation equations and thus provide a natural route to the Kerr problem. For these (and other) reasons, the time-domain approach to SF calculations has been favoured by a number of workers in the field, and is under active development by several groups \cite{Barack:Sago:2007, Vega:Diener:Tichy:Detweiler:2009,Thornburg:2009,Thornburg:2010,Canizares:Sopuerta:2009,Canizares:Sopuerta:Jaramillo:2010,Dolan:Barack:2011}. 

The obvious weakness of time-domain formulations lies in their computational costliness. The lack of separability of the perturbation equations into multipole harmonics on the Kerr background means that one has to deal with partial differential equations in 2+1D (or in full 3+1D). An added complexity is introduced by the presence of a pointlike source: the physical perturbation diverges along the worldline of a pointlike particle (the divergence is Coulomb-like in the 3+1D case and logarithmic in the 2+1D case \cite{Barack:Golbourn:2007}), 
and a naive numerical implementation would fail. One first has to ``regularize'' the field equations in order to make them amenable to numerical treatment. A regularization scheme formulated in 2+1D was introduced in Refs.\ \cite{Barack:Golbourn:2007,Barack:Golbourn:Sago:2007}, and a closely related method is being developed by Vega and collaborators \cite{Vega:Detweiler:2008,Vega:Diener:Tichy:Detweiler:2009,Vega:Wardell:Diener:2011} within a 3+1D framework. Both methods are based on the idea of a ``puncture'' function: Rather than evolving the physical (singular) field, one evolves a ``punctured'' version thereof, obtained by subtracting from the full field a certain (singular) puncture function, given analytically. In practice, this amounts to (numerically) solving the field equations with a certain ``effective source'' to obtain the residual, regularized field. If a suitable puncture function is chosen (see below), the physical SF can be constructed directly from the residual field. 

The 2+1D scheme of Refs.\ \cite{Barack:Golbourn:2007,Barack:Golbourn:Sago:2007} takes advantage of the axial symmetry of the Kerr background. The perturbation equations are separated into azimuthal ($\propto e^{im\varphi}$) modes and solved mode-by-mode, using an $m$-mode version of the puncture scheme. The SF is then reconstructed as a sum over $m$-mode contributions (see below for a more precise description). In Ref.\ \cite{Dolan:Barack:2011} (hereafter `Paper I') we presented a first full implementation of this {\it $m$-mode regularization scheme}, as applied to the scalar-field SF on a Schwarzschild background. We used this simplified model as a convenient platform for development, and in order to illustrate the generic properties of the method in a relatively ``clean'' environment. 

In the current work we take an important step forward by applying our method to the Kerr case. We thus present the first time-domain SF calculation in Kerr geometry. As in paper I, we consider a scalar-field toy model, and specialize to circular orbits (in the equatorial plane of the Kerr black hole). This lays the groundwork 
for a gravitational SF implementation in Kerr, which we will present in the third paper of the series \cite{in_progress}. 

In the remainder of this introductory section we review the $m$-mode regularization method in some detail, and give a short preview of the current work.

\subsection{$m$-mode regularization\label{subsec:m-mode-reg}}

Below is a schematic description of the $m$-mode regularization method as applied to the scalar-field SF in an axisymmetric spacetime such as Kerr (the basic structure of the scheme remains unaltered in the gravitational case). We model the compact object as a point particle carrying a scalar charge $q$ and moving along a geodesic of the background (e.g., Kerr) geometry. In our model, the scalar field $\Phi$ sourced by the particle is taken to satisfy the minimally-coupled Klein Gordon equation,
\beq \label{KG}
\Box\Phi=S ,
\eeq
where the covariant derivatives in $\Box\equiv \nabla^{\alpha}\nabla_{\alpha}$ (as elsewhere in this work) are taken with respect to the background (e.g.\ Kerr) metric $g_{\alpha\beta}$, and the source $S$ is that associated with the energy-momentum of the pointlike charge $q$ [modelled by a delta-function distribution; cf.\ Eq.\ (\ref{S-3d}) below]. In our notation, $\Phi$ represents the ``physical'', retarded solution to Eq.\ (\ref{KG}) (which is, of course, divergent at the particle's location). Let us recall that the physical SF acting on the scalar charge can be derived by means of the Detweiler--Whiting formula \cite{Detweiler:Whiting:2003}
\beq \label{Fself}
F^{\alpha}_{\rm self}=q\nabla^{\alpha}\Phi_R=q\nabla^{\alpha}(\Phi-\Phi_S).
\eeq
Here $\Phi_S$ (``S field'') is a specific singular solution of the inhomogeneous equation (\ref{KG}), and the difference $\Phi_R\equiv \Phi-\Phi_S$ (``R field'') is a smooth, homogeneous solution of the sourceless field equation, $\Box\Phi_R=0$. The formal construction of $\Phi_S$ is described in \cite{Detweiler:Whiting:2003}.

In most cases of interest, the singular field $\Phi_S(x)$ cannot be evaluated in exact form. However, the form of $\Phi_S(x)$ near the particle's worldline can be studied analytically to inform a local approximation $\Phipunc(x)$---the ``puncture function''---which may be defined globally through some extension far from the worldline. The ``order'' of the puncture describes how well it approximates $\Phi_S(x)$ near the particle. In Paper I we introduced the following convention: a puncture is said to be of ``order $n$'', denoted $\Phipunc^{[n]}$, if the difference $\Phipunc^{[n]}-\Phi_S$ is a $C^{n-2}$ function on the particle's worldline. In addition, we require $\Phipunc^{[n]}=\Phi_S$ as  well as $\nabla^{\alpha}\Phipunc^{[n]}=\nabla^{\alpha}\Phi_S$ at the particle limit (noting that the first of these conditions makes sense only for $n\geq 2$, and the second only for $n\geq 3$). Since the R field $\Phi_R$ is smooth ($C^{\infty}$), the {\it residual} field associated with the $n$-th order puncture, 
\beq
\Phires^{[n]}\equiv \Phi-\Phipunc^{[n]} = \Phi_R-(\Phipunc^{[n]}-\Phi_S) ,   \label{eq-residual}
\eeq
is a $C^{n-2}$ function on the worldline. Hence, for example, a 2nd-order residual function $\Phires^{[2]}$ is continuous but not differentiable, and a 3rd-order residual function $\Phires^{[3]}$ is both continuous and differentiable. For $n\geq 3$ we have $\nabla^{\alpha}\Phires^{[n]}=\nabla^{\alpha}\Phi_R$ at the particle limit, which allows us to write
\beq\label{Fself2}
F^{\alpha}_{\rm self}=q\nabla^{\alpha}\Phires^{[n\geq 3]}.
\eeq

Thus, the problem of calculating the SF reduces to the problem of computing the residual field $\Phires^{[n]}$ (for some $n\geq 3$) near the particle. This field satisfies a ``regularized'' version of the Klein-Gordon equation:
\beq \label{KGreg}
\Box\Phires^{[n]} = S(x) - \Box \Phipunc^{[n]} \equiv \Seff^{[n]}.   
\eeq
The {\em effective source} $\Seff^{[n]}(x)$ is an extended source that no longer involves a delta function; it is generally $C^{n-4}$ on the worldline. Hence, a 4th-order puncture gives rise to a continuous effective source, in general. Following the development of efficient computational tools for constructing high-order Green function expansions \cite{Ottewill:Wardell,Vega:Wardell:Diener:2011,Wardell:thesis}, a 4th-order puncture for generic orbits in Kerr geometry is now at hand, and higher-order extensions are possible.

The 3+1D treatment of Vega {\it et al.}\ \cite{Vega:Detweiler:2008,Vega:Diener:Tichy:Detweiler:2009,Vega:Wardell:Diener:2011} tackles the field equation (\ref{KGreg}) directly. In the 2+1D formulation, instead, we take advantage of the azimuthal symmetry of the background in order to reduce the dimensionality of the problem. Let $\varphi$ be a suitable ``azimuthal'' coordinate associated with the axial symmetry of the background (below we will make a specific choice for $\varphi$ in the Kerr case; note that the Boyer--Lindquist coordinate $\phi$ does not suite our purpose as it is singular at the event horizon of the Kerr black hole). We formally decompose the residual field into azimuthal modes in the form 
\beq
\Phires=\sum_{m=-\infty}^{\infty} \Phires^{m}e^{im\varphi},  \label{mmode-decomp}
\eeq
and similarly for $\Phipunc$ and $\Seff$ (we have omitted the labels $[n]$ for brevity). This decomposition separates the field equation (\ref{KGreg}), with each of the $m$-modes satisfying an equation of the form
\beq \label{KGregm}
\Box^m\Phires^{m} = \Seff^{m},
\eeq
where $\Box^m$ is a certain $m$-dependent differential operator in 2+1D [cf.\ Eq.\ (\ref{waveeq-3}) below]. Given the modes  $\Phires^{m}(x)$ near the particle, the SF can be reconstructed as a sum over modal contributions,
\beq \label{FselfSum}
F^{\alpha}_{\rm self}=\sum_{m=-\infty}^{\infty} F^{\alpha m}_{\rm self},
\quad\quad 
F^{\alpha m}_{\rm self}\equiv q\nabla^{\alpha}\left(\Phires^{m} e^{im\varphi}\right),
\eeq
evaluated at the particle. That the gradient indeed commutes with the modal sum (for $n\geq 2$) was shown in Ref.\ \cite{Barack:Golbourn:Sago:2007}. Ref.\ \cite{Barack:Golbourn:Sago:2007} also argued that the reconstruction formula (\ref{FselfSum}) applies, in fact, for any $n\geq 2$, even though the original 3+1D formula (\ref{Fself}) makes sense only for $n\geq 3$. Note that the individual modal contributions $F^{\alpha m}_{\rm self}$ depend on the puncture order $n$; however, the total SF $F^{\alpha}_{\rm self}$ is, of course, $n$-independent (for $n\geq 2$). 

In our method, the fields $\Phires^{m}(x)$ are obtained by solving Eq.\ (\ref{KGregm}) numerically for each $m$ using a finite-difference algorithm on a 2+1D mesh. The boundary conditions for $\Phires^{m}$ depend, in principle, on the global extension of the puncture $\Phipunc$. It is convenient to truncate the support of $\Phipunc$ far from the particle, in which case the residual $\Phires$ (and its modes $\Phires^{m}$) satisfy the usual ``retarded'' boundary conditions (outgoing radiation at infinity and purely ingoing radiation across the horizon; boundary conditions for any nonradiative modes are determined from regularity). This truncation is achieved in our scheme by introducing an auxiliary worldtube around the particle's worldline (in the 2+1D space), as we shall describe in Sec.~\ref{sec:worldtube}. 

In Paper I we implemented the above ``$m$-mode regularization scheme'' to compute the SF acting on a scalar charge in a circular orbit around a Schwarzschild black hole (refraining from a multipole decomposition and working instead in 2+1D). We implemented our numerical code with punctures of second, third, and fourth orders. In all cases our numerical results were found to be in good agreement with those of previous calculations using frequency-domain methods or time-domain integration in 1+1D, thus reaffirming the validity of the reconstruction formula (\ref{FselfSum}) for $n=2$--$4$. 

We used the simple Schwarzschild model in Paper I to explore some of the generic features of the $m$-mode method. The question of the convergence rate of the modal sum in Eq.\ (\ref{FselfSum}) is a crucial one, since in actual numerical implementations one can only compute a finite number of modes (realistically, no more than $\sim 20$ modes using our current code). We made the following empirical observations (supported by heuristic arguments): (i) The {\it dissipative} piece of the SF has an exponentially convergent mode sum; (ii) the {\it conservative} piece of the SF has a mode-sum that converges with an inverse-power law: the individual modal contributions in the mode sum fall off at large $m$ as $m^{-n}$ for even $n$ and as $m^{-n+1}$ for odd $n$ (where $n$ is the order of the puncture used). Thus, the large-$m$ modal contribution to the conservative piece of $F^{\alpha}_{\rm self}$ is $\propto m^{-2}$ with a 3rd-order puncture, and $\propto m^{-4}$ with either a 4th or a 5th-order puncture. We concluded that a 4th-order puncture implementation offers an optimal balance between simplicity (of the analytic puncture formulation) and efficiency (of the computational algorithm). We expect this conclusion to carry over to the Kerr case.

In our implementation (as in any other time-domain implementation) the initial data for the time evolution are, of course, unknown. Instead, we begin the evolution with arbitrarily selected initial data, and rely on the property of homogeneous perturbations in black hole geometries to completely ``dissipate away'' at late time (``no hair theorem''). We let the evolution proceed until the solution relaxes to a stationary state at a sufficient level. It is important to understand the rate of relaxation, since this determines the necessary evolution time and thus has important effect on the computational burden. As discussed in Paper I, the theoretical expectation is for the modal contributions to relax with an $m$-dependent inverse-power law in time $t$. Modes with higher $m$ relax faster, and the slowest relaxation is exhibited by the $m=0$ mode: $\propto t^{-2}$ for the field modes $\Phires^{m}$ and $\propto t^{-3}$ for the force modes $F^{\alpha m}_{\rm self}$. This behavior was confirmed numerically in Paper I; we expect it, too, to carry over to the Kerr case. The issue points to an important advantage of the 2+1D formulation (as compared with the full 3+1D treatment): Since different $m$-modes are evolved separately, we have the flexibility of being able to adjust the evolution time as a function of $m$ to achieve computational saving.

\subsection{This paper}

Here we extend the analysis of Paper I to the Kerr case, focusing again on circular motion (in the equatorial plane). Our 2+1D treatment makes such a generalization straightforward {\em in principle}. However, there are several nontrivial technical details that require careful consideration and extra development. First, the analytic expressions involved in the puncture formulation become considerably more complicated and require commensurably greater care in their numerical implementation. Second, the azimuthal coordinate used to define the $m$-mode decomposition must be chosen judiciously; the standard Boyer-Lindquist coordinate suits for purpose in Schwarzschild geometry but not in the Kerr case. The third point is most crucial: We find that the finite-difference scheme employed in Paper I to evolve the field equation (\ref{KGregm}) becomes unstable in the Kerr case and must be replaced with an alternative method. We have experimented with several alternatives, and will report here our findings. We observe that some of these methods, used previously in the literature to study vacuum perturbations on Kerr, develop instabilities at large values of $m$, which makes them unsuitable for us. We develop and implement a finite-difference scheme (a variant of the ``method of lines'') that appears to perform well even for large $m$ values. Our scheme is second-order convergent (in the grid spacing),
and we implement it here with a puncture of 4th order ($n=4$).

The remainder of this paper is structured as follows. In Sec.~\ref{sec:scalar-field} we describe our physical setup and review the formalism of scalar-field perturbations on a Kerr background. In Sec.~\ref{sec:punc_and_source} we formulate the $m$-mode regularization scheme as applied to the problem at hand, and prescribe 2nd, 3rd and 4th-order puncture functions with corresponding effective sources. Section \ref{sec:numerical_method} describes our numerical method, including the worldtube technique and the finite-difference scheme. We also report on some of our less-successful experiments with other schemes. In Section \ref{sec:results} we present some numerical results along with convergence tests, and explore the late-time relaxation of the numerical solutions and the convergence properties of the $m$-mode sum. We demonstrate that our code successfully reproduces the results of Warburton--Barack's frequency domain calculation \cite{Warburton:Barack:2010}. In the concluding section, Sec.\ \ref{outlook}, we discuss future directions, including the application of our method to the gravitational SF. 

Throughout this work we use metric signature ${-}{+}{+}{+}$ and set $G=c=1$.

\section{Preliminaries}\label{sec:scalar-field}

\subsection{Setup and notation}

In Boyer-Lindquist coordinates $(t,r,\theta,\phi)$, the line element $ds^2 = g_{\mu \nu} dx^{\mu} dx^{\nu}$ of a Kerr spacetime with mass $M$ and spin parameter $a$ is given by
\begin{equation*}
ds^2 = -\left(1 - \frac{2Mr}{\rho^2}\right) dt^2 - \frac{4aMr\sin^2\theta}{\rho^2} dt d\phi + \frac{\rho^2}{\Delta} dr^2 + \rho^2d\theta^2 + \left( r^2+a^2+\frac{2Mr a^2 \sin^2\theta}{\rho^2} \right)\sin^2\theta d\phi^2 ,
\end{equation*}
where
\beq
\rho^2 = r^2 + a^2 \cos^2 \theta \quad \quad \text{and} \quad \quad \Delta = r^2 - 2Mr + a^2 .   \label{rho-Delta-def}
\eeq
The spacetime admits two Killing vectors, $\xi^{\mu}_{(t)} = \delta^{\mu}_{t}$ and $\xi^{\mu}_{(\phi)} = \delta^{\mu}_\phi$, as well as a Killing tensor $Q^{\mu\nu}$. The event horizon is the hypersurface $r=r_+$, where $r_+$ is the larger of the two roots of $\Delta(r)$, given by $r_{\pm} = M \pm \sqrt{M^2 - a^2}$. The ergosphere (inside which static observers do not exist) extends between the event horizon and the static limit surface $r_{sl} = M + \sqrt{M^2 - a^2 \cos^2 \theta}$. Stationary observers just outside the event horizon have angular velocity ($d\phi/dt$) given by
\beq
\Omega_H=\frac{a}{2Mr_+},
\eeq 
which may be interpreted as the angular velocity of the horizon. Timelike geodesics of the Kerr geometry can be parametrized by the three constants of motion
\beq
\EE = - \xi^{\mu}_{(t)} u_\mu, \quad\quad \LL = \xi^{\mu}_{(\phi)} u_\mu, \quad \quad  \mathcal{Q} = Q^{\mu \nu} u_{\mu} u_{\nu} ,
\eeq
referred to as the (specific) energy, azimuthal angular momentum and Carter constant, respectively. In the last expressions $u_{\mu}=g_{\mu\nu}u^{\nu}$, where $u^{\nu}$ is the tangent four-velocity along the geodesic, and, as elsewhere in this work, we use the background (Kerr) metric to raise and lower indices.

In this work we consider the scalar-field SF on a pointlike scalar charge moving on a circular geodesic orbit in the equatorial plane of the central Kerr black hole. The motion may be either prograde (i.e., in the same sense as the rotation of the black hole) or retrograde. We shall adopt the convention that $u^{\phi}$ (hence also ${\cal L}_z$) is always positive, and distinguish between prograde and retrograde orbits by the sign of $a$: $a>0$ for the former; $a<0$ for the latter. 

For geodesics in the equatorial plane ($\theta=\pi/2$) one finds ${\cal Q}=0$. If the orbit is both equatorial and circular, the other two constants of motion are given by 
\begin{equation}
\mathcal{E} = \frac{1 - 2v^2 + \tilde{a}v^3}{\sqrt{1 - 3v^2 + 2\tilde{a}v^3}},    
\quad\quad 
\mathcal{L}_z =r_0 v\, \frac{1-2\tilde{a}v^3 + \tilde{a}^2v^4}{\sqrt{1-3v^2+2\tilde{a}v^3}} ,				
\end{equation}
where $r_0$ is the Boyer-Lindquist radius of the orbit, $v\equiv\sqrt{M/r_0}$ and $\tilde{a}\equiv a/M$.
The particle's angular frequency is given by
\begin{equation}
\Omega\equiv \frac{d\phi}{dt} =\frac{u^{\phi}}{u^t}= 
\frac{v^3}{M(1+\tilde a v^3)}.
\end{equation}

\subsection{The scalar-field equation in 2+1D}

The scalar field is assumed to be governed by the sourced (minimally coupled) Klein-Gordon equation, 
\beq \label{waveeq}
\Box \Phi(x) = \frac{1}{\sqrt{-g}}\, \partial_\mu \left[ \sqrt{-g}\, g^{\mu \nu} \partial_\nu \Phi(x) \right] = S(x) ,
\eeq
where $g = -\rho^4 \sin^2 \theta$ is the Kerr metric determinant, and $g^{\mu \nu}$ is the contravariant form of the metric. The source $S(x)$ is taken to have a delta-function support on the geodesic worldline: 
\beq
S(x) = -4 \pi q \int_{-\infty}^{\infty} \left[ -g(x)\right]^{-1/2} \delta^4\left[ x - \bar x(\tau') \right] d\tau' .  \label{S-3d}
\eeq 
Here $q$ is the scalar charge of the particle, $\tau$ is its proper time, and $x=\bar x(\tau)$ [shorthand for $x^{\alpha}=\bar x^{\alpha}(\tau)$] parameterizes the particle's geodesic orbit. In our setup we have
$\bar r=r_0$(=const) and $\bar\theta=\pi/2$, as well as $\bar\phi=\Omega\, t$ (taking $\bar\phi=0$ at $t=0$ without loss of generality).

We now wish to decompose the field equation (\ref{waveeq}) into azimuthal modes. We must be mindful of the fact that the Boyer-Lindquist azimuthal coordinate $\phi$ is pathological at the event horizon. We cure this by making the standard coordinate change $\phi\to\varphi$ defined through
\begin{equation}
d\varphi = d \phi + \frac{a}{\Delta} dr  ,
\end{equation}
which, choosing the constant of integration so that $\varphi$ coincides with $\phi$ as $r\to\infty$, gives
\beq
\varphi(\phi,r) =\phi + \frac{a}{r_+ - r_-} \ln \left| \frac{r - r_+}{r - r_-} \right| . \label{Delta-phi}
\eeq
Since both $\Phi$ and $S$ are periodic in $\varphi$, they admit $m$-mode decompositions (formally, Fourier expansions) of the form 
\beq
\Phi=\sum_{m=-\infty}^{\infty} \Phi^m e^{im\varphi}, \quad\quad
S=\sum_{m=-\infty}^{\infty} S^m e^{im\varphi}.
\eeq
 The $m$-modes $\Phi^m$ may be obtained via the inversion formula
\beq
\Phi^m(t,r,\theta) = \frac{1}{2\pi} \int_{-\pi}^{\pi} e^{-i m {\varphi}} \Phi(t,r,\theta,\varphi) d \varphi,  \label{eq:inversion}
\eeq
and similarly for $S^m$. For the latter we obtain, performing the integration,
\beq
S^m= -\frac{2q}{r_0^2 u^t} \delta(r-r_0)\delta(\theta-\pi/2)e^{-im\bar\varphi(t)},  \label{S-2+1d}
\eeq
where $\bar\varphi(t)\equiv\varphi(\bar\phi(t),r_0)$. Here we used the notation $\bar\phi(t)\equiv \bar\phi(\tau(t))$, where the function $\tau(\bar t)$ is obtained by formally inverting the (monotonically increasing) function $\bar t(\tau)$.


The above decomposition separates Eq.\ (\ref{waveeq}) into a set of 2+1D equations for the individual (complex-valued) $m$-modes $\Phi^m(t,r,\theta)$. To write these equations in a form more amenable to numerical treatment via the Method of Lines (see Sec.~\ref{subsec:FDmethod} below), we introduce two minor modifications:
we transform to the new variable 
\beq
\Psim\equiv r\Phim,
\eeq
and we replace $\partial_r\to \partial_{r_*}$, where the Kerr tortoise coordinate $r_\ast$ is defined (as usual) by
\beq
\frac{d r_\ast}{d r} = \frac{r^2 + a^2}{\Delta} , 
\eeq
or, specifying the constant of integration,
\beq
r_\ast = r + \frac{2M}{r_+ - r_-} \left( r_+ \ln \left| \frac{r - r_+}{2M} \right| - r_- \ln \left| \frac{r-r_-}{2M} \right|  \right)  .
\eeq
This brings the separated wave equation into the final, working form
\beq \label{Boxm}
\Box^m_{\Psi}\Psim = S^m_{\Psi}, 
\eeq
where the $m$-mode scalar wave operator reads
\begin{eqnarray}
\Box^m_{\Psi}&\equiv& \frac{\partial^2 }{\partial t^2} 
+ \frac{4 i a m M r}{\Sigma^2} \frac{\partial }{\partial t}
- \frac{(r^2+a^2)^2}{\Sigma^2} \frac{\partial^2 }{\partial r_\ast^2} 
-\left[\frac{2iamr(r^2+a^2)-2a^2\Delta}{r\Sigma^2}\right]\frac{\partial }{\partial r_\ast} \nonumber \\
&& - \frac{\Delta}{\Sigma^2} \left[   \frac{\partial^2}{\partial \theta^2} + \cot\theta \frac{\partial}{\partial \theta} 
- \frac{m^2}{\sin^2 \theta} - \frac{2M}{r}\left(1-\frac{a^2}{Mr}\right) - \frac{2iam}{r}
 \right]  ,  \label{waveeq-3}
\end{eqnarray}
the $m$-mode source is given by 
\beq
S^m_{\Psi}= - \frac{r \Delta \rho^2}{\Sigma^2} S^m,  \label{mmode-source}
\eeq
and we have introduced
\beq
\Sigma^2 \equiv   (r^2 + a^2)^2 - a^2 \Delta \sin^2 \theta .
\eeq

\section{Puncture formulation \label{sec:punc_and_source}}

The retarded solutions of the $m$-mode field equation (\ref{Boxm}) diverge on the particle's worldline for all $m$. [The divergence is logarithmic in the spatial distance to the particle, and its leading-order form does not depend on $m$---cf.\ Eq.\ (23) of Ref.\ \cite{Barack:Golbourn:2007}.] In our approach we do not tackle Eq.\ (\ref{Boxm}) directly, but rather consider a ``punctured'' version thereof, given by 
\beq \label{KGregPsim}
\Box^m_{\Psi}\Psires^{[n]m} = S_{\Psi{\rm eff}}^{[n]m}.
\eeq
Here the variables $\Psires^{[n]m}$ are defined as the $m$-modes of the $n$th-order residual field 
\beq
\Psires^{[n]}=r(\Phi-\Phipunc^{[n]})=r\Phires^{[n]},
\eeq
with $S_{\Psi{\rm eff}}^{[n]m}$ being the $m$-modes of the associated $n$th-order effective source
\beq\label{SeffPsi}
S_{\Psi{\rm eff}}^{[n]} = - \frac{r\Delta\rho^2}{\Sigma^2}\left(S-\Box\Phipunc^{[n]}\right).
\eeq
With a suitable puncture $\Phipunc^{[n\geq 2]}$, the physical SF is then constructed via
\beq \label{FselfSum2}
F^{\alpha}_{\rm self}=\sum_{m=-\infty}^{\infty} F^{\alpha m}_{\rm self},
\quad\quad 
F^{\alpha m}_{\rm self}\equiv q\nabla^{\alpha}\left(r^{-1}\Psires^{m[n]} e^{im\varphi}\right),
\eeq
evaluated at $\bar x(\tau)$. To complete the formulation one need only prescribe a suitable puncture $\Phipunc^{[n]}$, which can then be used to construct (analytically) the effective source $S_{\Psi{\rm eff}}^{[n]}$ via Eq.\ (\ref{SeffPsi}). 

In what follows we construct puncture functions $\Phipunc^{[n]}$ of 2nd, 3rd and 4th orders, based a local expansion of the Detweiler--Whiting S field. Similar punctures have been developed in previous work \cite{Haas:Poisson:2006, Barack:Golbourn:Sago:2007,Vega:Wardell:Diener:2011}, but we give them here (and in Appendix \ref{appendix:4thord}) in explicit ready-to-use coordinate form, specialized to the problem at hand of circular equatorial orbits in Kerr. We then also describe the procedure for constructing the $m$ modes $\Phipunc^{[n]m}$ and $S_{\Psi{\rm eff}}^{[n]m}$, and give explicit analytic expressions for $n=2$.

\subsection{Punctures through fourth order\label{subsec:punc}}
Our first task is to obtain a local expansion for the Detweiler--Whiting S field $\Phi_S(x)$ in the vicinity of a point on the worldline, $\bar x$. Let us presume that $x$ and $\bar x$ are connected by a unique spacelike geodesic, so that we may introduce the Synge bi-scalar $\sigma(x,\bar{x})$ \cite{Synge:1960,Poisson:Pound:Vega:2011}, which is equal to one half of the squared geodesic distance along the geodesic connecting $x$ and $\bar{x}$. Furthermore, we presume that there exists a sufficiently-regular coordinate chart which covers a region of spacetime encompassing $x$, $\bar{x}$ and the spacelike geodesic. In this chart, points $x$ and $\bar{x}$ take coordinates $x^\alpha$ and $\bar{x}^\alpha$, and the vector tangent to the worldline at $\bar x$ is denoted by $\bar u^{\alpha}$. Now we let $\bar\sigma_{\alpha}$ denote the covariant derivative of $\sigma$ taken with respect to $\bar x$, and we introduce the orthogonal and parallel projections
\beq \label{epsilon}
\epsilon^2 \equiv (\bar g^{\alpha\beta}+\bar u^{\alpha} \bar u^{\beta})\bar\sigma_{\alpha} \bar\sigma_{\beta},
\quad\quad
\eta \equiv \bar u^{\alpha} \bar\sigma_{\alpha} ,
\eeq
where $\bar g^{\alpha\beta}$ is the background metric evaluated at $\bar x$. Note that $\epsilon$ has the interpretation of the spatial geodesic distance from the field point $x$ to the worldline, i.e., the length of the spatial geodesic segment connecting $x$ to the worldline and normal to it \footnote{In Refs.~\cite{Haas:Poisson:2006,Vega:Wardell:Diener:2011} the symbol $s$ was used in the place of our $\epsilon$.}.

At this point we introduce the coordinate difference $\delta x$, with coordinates
\beq
\delta x^{\alpha} =  x^\alpha - \bar x^\alpha .
\eeq
Our aim below is to obtain a coordinate expansion for the Detweiler--Whiting S field, $\Phi_S(x^\alpha)$, in powers of $\delta x^{\alpha}$. 
When $\delta x^\alpha$ is sufficiently small, $\Phi_S$ can be shown to admit the local expansion \cite{Haas:Poisson:2006,Vega:Wardell:Diener:2011}
\begin{eqnarray}\label{Sfield}
\Phi_S(x) =\frac{q}{\epsilon}\left[1 + \frac{\alpha(x,\bar x)}{\epsilon^2} +\frac{\beta(x,\bar x)}{\epsilon^2} + O(\delta x^3) \right],
\end{eqnarray}
where 
\begin{equation}\label{alpha}
\alpha=\frac{1}{6}\left(\eta^2 - \epsilon^2\right)R_{\alpha\beta\gamma\delta}(\bar x) \bar u^{\alpha} \bar u^{\gamma} \bar\sigma^{\beta} \bar\sigma^{\delta},
\end{equation}
and
\begin{equation}\label{beta}
\beta=\frac{1}{24}\Big[\eta(\eta^2 - 3\epsilon^2) \nabla_{\mu}R_{\alpha\beta\gamma\delta}(\bar x) \bar u^{\alpha} \bar u^{\gamma} \bar u^{\mu} \bar\sigma^{\beta} \bar\sigma^{\delta} - (\eta^2 - \epsilon^2) \nabla_{\mu} R_{\alpha\beta\gamma\delta}(\bar x) \bar u^{\alpha} \bar u^{\gamma} \bar\sigma^{\beta} \bar\sigma^{\delta} \bar\sigma^{\mu} \Big] .
\end{equation}
Here $R_{\alpha\beta\gamma\delta}$ and $\nabla_{\mu}R_{\alpha\beta\gamma\delta}$ are the background (Kerr) Riemann tensor and its covariant derivative (both evaluated here at $\bar{x}$), and $\bar\sigma^{\alpha}=\bar g^{\alpha\beta}\bar\sigma_{\beta}$. Note the bi-scalars $\alpha$ and $\beta$ are smooth ($C^{\infty}$) functions of $x$ and $\bar x$ (and hence $\delta x$), with $\alpha=O(\delta x^4)$ and $\beta=O(\delta x^5)$ near the worldline. Hence, in Eq.\ (\ref{Sfield}) the terms $\alpha/\epsilon^3$ and $\beta/\epsilon^3$ are, respectively, $C^0$ and $C^1$ functions of $\delta x$ at the worldline. Therefore, defining the ``truncated'' S fields 
\begin{equation}
\Phi_S^{[2]}=q/\epsilon, \quad\quad
\Phi_S^{[3]}=q(\epsilon^{-1}+\alpha\epsilon^{-3}),\quad\quad
\Phi_S^{[4]}=q(\epsilon^{-1}+\alpha\epsilon^{-3}+\beta\epsilon^{-3}),
\end{equation}
we find that the differences $\Phi_S-\Phi_S^{[n]}$ are $C^{n-2}$ at the worldline.

The local behavior of the functions $\Phi_S^{[2,3,4]}$ near the worldline is consistent with that of punctures $\Phipunc^{[2,3,4]}$ (correspondingly). However, in this form $\Phi_S^{[n]}(x)$ cannot be used in a practical scheme for two reasons: (i) $\bar \sigma_\alpha$ [and therefore $\Phi_S^{[n]}(x)$] is not known in explicit analytic form 
and (ii) $\bar \sigma_\alpha$ [and therefore $\Phi_S^{[n]}(x)$] is only defined if the field point $x$ is in the vicinity of the worldline point $\bar x$. To construct globally-defined analytically-given punctures we take two steps: (i) We expand $\epsilon^2$, $\alpha$ and $\beta$ in powers of $\delta x$, to obtain a local coordinate expansion of $\Phi_S^{[n]}$ up to a suitable order. (ii) We promote the local expansion to a global function by choosing an appropriate analytic extension. 

To take step (i), we use the method of Ref.\ \cite{Ottewill:Wardell} to obtain a Taylor expansion for $\bar\sigma_{\alpha}$ in powers of $\delta x$ (with fixed $\bar{x}$). 
We use the Taylor expansion in Eqs.\ (\ref{epsilon}), (\ref{alpha}) and (\ref{beta}) to obtain expansions for the 
 bi-scalars $\epsilon^2$, $\alpha$ and $\beta$. Let us use the notation $\esq_{(n)}$, $\alpha_{(n)}$ and $\beta_{(n)}$ to denote the Taylor expansion of the corresponding quantities up to and including the $O(\delta x^n)$ term.

To take step (ii), we allow the point on the worldline $\bar{x}$ to become a function of the field point $x$, i.e., $\bar{x}(x)$, in such a way that $x$ and $\bar{x}(x)$ are connected by a unique spatial geodesic as $x$ approaches the worldline. A simple way to achieve this is to take $\bar{t} = t$, where from here on we specialize to Boyer-Lindquist coordinates. Namely, we relate a field point $x^{\alpha} = (t, r, \theta, \phi)$ to a worldline point $\bar{x}^\alpha = (t, r_0, \pi/2, \Omega t)$. Then the coordinate differences are simply $\delta t = 0$, $\delta r = r - r_0$ and $\delta \theta = \theta - \pi /2$. We note that $\phi - \Omega t$ is not periodic, and it may be large even when the points are close together. To rectify this we instead choose $\delta\phi=2\sin^{-1}\left\{\sin[(\phi-\Omega t)/2]\right\}$ where the principal values $-\pi/2\leq \sin^{-1} x\leq \pi/2$ are assumed. 

After taking steps (i) and (ii), we obtain truncated Taylor series coordinate expansions $\esq_{(n)}$, $\alpha_{(n)}$ and $\beta_{(n)}$ in place of the geometric quantities $\epsilon^2$, $\alpha$ and $\beta$. Explicitly, $\epsilon^2 = \esq_{(n)} + O(\delta x^{n+1})$, where
\begin{equation}\label{esqn}
\esq_{(n)}  = M^2 \sum\limits_{\genfrac{}{}{0pt}{}{i,j,k=0}{2 \le i+j+k\le n}}^n {\sco}_{ijk}(\delta \tilde r)^i(\delta\theta)^j(\delta\phi)^k ,   
\end{equation}
and likewise for $\alpha$ and $\beta$. 
Here, for convenience, we have introduced the dimensionless radial distance $\delta\tilde r\equiv (r-r_0)/M$, and ${\sco}_{ijk}$ are dimensionless coefficients depending only on $r_0$ and $a$. 
In Appendix \ref{appendix:4thord} we give explicit expressions (for the specific case of circular equatorial geodesics on Kerr spacetime) for all coefficients ${\sco}_{ijk}$ necessary for computing $\esq_{(n)}$ through $n=5$. We also give explicit expressions for $\alpha_{(n)}$ and $\beta_{(n)}$ through $n=5$.


With the quantities $\esq_{(3,4,5)}$, $\alpha_{(4,5)}$ and $\beta_{(5)}$ all at hand in explicit analytic form, 2nd, 3rd and 4th-order punctures can be prescribed as follows:
\begin{equation}\label{eq:punc-2nd}
\Phipunc^{[2]} = \frac{q}{\esq^{1/2}_{(3)}},
\end{equation}
\beq \label{eq:punc-3rd}
\Phipunc^{[3]} =q\left(\frac{1}{\esq^{1/2}_{(4)}} + \frac{\alpha_{(4)}}{\esq^{3/2}_{(2)}}\right),
\eeq
\begin{equation} \label{eq:punc-4th}
\Phipunc^{[4]} = q \left(\frac{1}{\esq^{1/2}_{(5)}} + \frac{\alpha_{(5)}}{\esq_{(3)}^{3/2}} + \frac{\beta_{(5)}}{\esq_{(2)}^{3/2}}\right).
\end{equation}
Note that in these expressions we have {\it relaxed} the assumption that $\delta x^\alpha$ is small in any sense, and allow $x$ to be any field point in the spacetime outside the black hole. (In practice we shall only need $x$ to lie within a certain finite-width worldtube surrounding the worldline---see Sec.\ \ref{sec:worldtube} below.) Since we work in a particular coordinate system (i.e., Boyer--Lindquist), this procedure amounts to choosing a particular global extension of $\Phi_S$ off the worldline. The above construction guarantees that our globally prescribed punctures $\Phi_{\cal P}^{[n]}$ approximate the S field $\Phi_S$ at the required level near the particle's worldline, i.e., that 
$\Phi_{\cal P}^{[n]}=\Phi_S$ up to a $C^{n-2}$ function.

\subsection{$m$-mode decomposition of the puncture field}\label{sec:m-mode}

We now describe the construction of the puncture $m$-modes $\Phipunc^{[n]m}$ for $n=2$ and $n=4$. We skip the $n=3$ case, which will not be considered further in this work. 

\subsubsection{Second-order puncture} 
The method for obtaining the $m$-mode decomposition of a 2nd-order puncture field of the form (\ref{eq:punc-2nd}) was described in Paper I. The first step is to recast $\Phipunc^{[2]}$ as a periodic function of $\delta\phi$, without any loss of order $n$. For $n=2$, this may be achieved by making the replacement
\beq
\delta\phi^2 \rightarrow 2 (1 - \cos \delta\phi ) = \delta\phi^2 + O\left(\delta\phi^4\right).
\eeq
Note that the puncture function involves only even powers of $\delta\phi$, and so it suffices to replace $\delta\phi^2$. 
Next, following the steps of Paper I [and using Eq.~(\ref{eq:inversion})] we write
\begin{eqnarray} \label{eq:m-decompose1}
\Phipunc^{[2]m} = \frac{1}{2\pi} \int_{-\pi}^{\pi} e^{-im\varphi} \Phipunc^{[2]} d \varphi = \frac{e^{-im\Delta \phi}}{2\pi} \int_{-\pi}^{\pi} e^{-im\phi} \Phipunc^{[2]} d \phi   ,
\end{eqnarray}
where we have introduced
\beq
\Delta \phi \equiv \varphi - \phi =  \frac{a}{r_+ - r_-} \ln \left| \frac{r - r_+}{r - r_-} \right| \label{delphi-def}
\eeq
[recall Eq.\ (\ref{Delta-phi})]. The integrand in Eq.\ (\ref{eq:m-decompose1}) has the form 
$(q/M) e^{-im\phi} \left[ A+2B(1-\cos\delta\phi)\right]^{-1/2}$, 
with 
\beq \label{AB}
A ={\sco}_{200} \delta\tilde r^2 + {\sco}_{020} \dth^2 + {\sco}_{300} \delta\tilde r^3 + {\sco}_{120} \delta\tilde r \dth^2,
\quad\quad
B = {\sco}_{002} + {\sco}_{102} \delta\tilde r,
\eeq
where the coefficients ${\sco}_{ijk}$ are those given in Appendix \ref{appendix:4thord}.
This integral can be expressed in terms of the complete elliptic integrals of the first and second kinds, respectively $K(k) \equiv \int_0^{\pi/2} \left(1 -  k^2 \sin^2 x \right)^{-1/2} dx$ and $E(k) \equiv \int_0^{\pi/2} \left(1 -  k^2 \sin^2 x \right)^{1/2} dx$, giving
\begin{eqnarray} \label{eq:m-decompose}
\Phipunc^{[2]m} =\frac{q e^{-i m (\Omega t + \Delta \phi )}}{2 \pi M B^{1/2}} \gamma \left[ p_K^m({\rhohat}) K(\gamma) + p_E^m({\rhohat}) E(\gamma) \right],
\end{eqnarray}
where
\beq
 {\rhohat} \equiv  [A / (4B)]^{1/2},  
 \quad \quad 
\gamma \equiv \left(1 + {\rhohat}^2 \right)^{-1/2}.   \label{rhohat-gamma-def}
\eeq
The quantities $p_K^m(\rhohat)$ and $p_E^m(\rhohat)$ are polynomials (of order $m$) in $\rhohat^2$; these polynomials are tabulated for $m=0$--$5$ in Table I of Ref.~\cite{Barack:Golbourn:2007} and can be readily obtained for higher values of $m$.

\subsubsection{Fourth-order puncture}
To recast $\Phipunc^{[4]}$ as a periodic function, without diminishing the order $n=4$ of the expansion, we made the replacement
\begin{align}
\delta\phi^2 \rightarrow& \; \frac{5}{2} - \frac{8}{3} \cos \delta\phi + \frac{1}{6} \cos 2 \delta\phi = \delta\phi^2 + O(\delta\phi^6) .  \label{deltaphi-periodic}  
\end{align}
Note that the 4th-order puncture, like the 2nd-order puncture, is symmetric under $\delta\phi \rightarrow -\delta\phi$, and hence it features only $\delta \phi^2$ and $\delta \phi^4$ terms. In our implementation we chose to replace $\delta \phi^4$ with the square of Eq.~(\ref{deltaphi-periodic}). An alternative choice would be to replace $\delta \phi^4$ with a linear function of harmonics $\cos(k \delta \phi)$ (where $k=0,1,2$) via
\begin{align}
\delta\phi^4 \rightarrow& \; 6 - 8 \cos \delta\phi + 2 \cos 2 \delta\phi = \delta\phi^4 + O(\delta\phi^6).
\end{align}

The $m$-mode decomposition proceeds 
as in Eq.~(\ref{eq:m-decompose1}). In this case, however, the form of the fourth-order puncture given in Eq.~(\ref{eq:punc-4th}) is such that the integration cannot be readily written in terms of elliptic integrals. 
Instead, we proceed by direct numerical integration of Eq.~(\ref{eq:m-decompose1}).

\subsection{Effective source}
The effective source $S^{[n]}_{\Psi{\rm eff}}$ is found by applying the wave operator $\Box$ to the puncture [see Eq.~(\ref{SeffPsi})], recalling that the dependence of $\Phipunc^{[n]}$ on $t$ and $\phi$ is only through the combination $\phi - \Omega t$ (for circular equatorial orbits). In practice, this implies that
\beq
\frac{\partial}{\partial t} = -\Omega \frac{\partial}{\partial ( \delta\phi )} , \quad \quad \frac{\partial}{\partial \phi} = \frac{\partial}{\partial (\delta\phi)} .
\eeq
The $m$-mode quantities $S^{[n]m}_{\Psi{\rm eff}}$ are then obtained via an inversion formula analogous to Eq.~(\ref{eq:inversion}). In the 2nd-order case, the integrals are readily computed analytically in terms of elliptic integrals (see Appendix \ref{appendix:2nd-ord}). In the 4th-order case, we found
it to be more straightforward to compute the integrals via direct numerical integration.

\section{Numerical Method\label{sec:numerical_method}}

In this section we describe the key features of our numerical implementation. These features are identical for any puncture order, so we shall omit here the order identifier `$[n]$' from our expressions, for brevity.

\subsection{Worldtube scheme and the numerical domain} \label{sec:worldtube}

Our puncture function $\Phi_{\cal P}$ and effective source $S_{\Psi{\rm eff}}$ are defined through a naive extension of a local expansion, and their behavior far from the particle may well prove problematic. Indeed, the 4th-order puncture in the form given in Eq.~(\ref{eq:punc-4th}) generally diverges at $r\to\infty$, and so does the effective source associated with this puncture. 
To handle this behaviour, we construct (following Ref.~\cite{Barack:Golbourn:2007} and Paper I) an auxiliary ``worldtube'' $\mathcal{T}$ in the 2+1D domain to enclose the worldline. Outside the worldtube, we evolve the finite-difference (FD) version of the homogeneous wave equation (\ref{Boxm}) for the full field modes $\Psi^m$. Inside the worldtube, we evolve the FD version of the sourced wave equation (\ref{KGregPsim}) for the residual field modes $\Psires^{m}$.  Across the boundary of the auxiliary tube, $\partial{\cal T}$, we convert between $\Psi^m$ and $\Psires^{m}$ using the  analytically-given value of the puncture field. To summarise,
\beq\label{tube}
\begin{cases} 
 \Box^m_{\Psi} \Psi^m = 0 , & \text{outside } \mathcal{T} , \\ 
 \Box^m_{\Psi} \Psires^{m} = S_{\Psi{\rm eff}}^m, & \text{inside } \mathcal{T} , \\
 \Psires^{m} = \Psi^m - r \Phipunc^{m}, & \text{across } \partial \mathcal{T} .
\end{cases}
\eeq
The full field modes $\Psi^m$, of course, satisfy the usual physical boundary conditions at the asymptotic domains (see below).

Our FD scheme is formulated on a fixed uniform grid based on $t,r_*,\theta$ coordinates; see Fig.~\ref{fig:grid}. The grid spacings in the corresponding coordinate directions are denoted $\Delt$, $\Delr$, $\Delq$. In the 2+1D domain, the particle's trajectory traces a straight line at $\theta=\pi/2$ and $r_*=r_{\ast0}[\equiv r_\ast(r_0)]$. For convenience, we lay the grid so that the particle's trajectory crosses through a row of grid points (of fixed $r_*,\theta$).

On this grid, we use a worldtube ${\cal T}$ of fixed coordinate widths $\{\Gamma_{r_\ast}, \Gamma_\theta \}$ centered at the worldline. Consider an arbitrary grid point with coordinates $(t,r_\ast, \theta)$. If $|r_\ast -  r_{\ast0}| \le \Gamma_{r_\ast} / 2$ and $|\theta - \pi/2| \le \Gamma_{\theta} / 2$ then the point is said to lie within the worldtube; otherwise it lies outside. For convenience, we choose $\Gamma_{r_\ast}$ and $\Gamma_\theta$ to be integer multiples of the grid spacings $\Delr$ and $\Delq$, respectively.


Within the FD scheme described below, the value of the field at a given grid point (say, point ``o'') is computed based on the values at several neighboring grid points, obtained in previous steps. If all these grid points lie either inside or outside $\cal T$, then the FD scheme is implemented straightforwardly. However, if point o lies close enough to $\partial{\cal T}$ that some of these points are ``in'' and others are ``out'', we make the following adjustment. If the point o is ``out'', then we first demote all ``in'' points to ``out'' points using $\Psi^m = \Psires^m + r\Phipunc^m$, before applying the FD formula. If, conversely, the point o is ``in'', then we promote all ``out'' points using $\Psires^m = \Psi^m - r\Phipunc^m$, before applying the FD procedure. In this way, we make use of the known value of $\Phipunc^m$ in order to communicate between the two numerical variables $\Psi^m$ and $\Psires^{m}$ across the boundary of the worldtube.


\subsection{Finite-difference scheme: Method of Lines\label{subsec:FDmethod}}

We describe our FD scheme as applied either inside or outside $\cal T$. For concreteness, we refer specifically to the full field equation (\ref{Boxm}), but note that the same procedure is applied inside $\cal T$, with the replacements $\Psi^m\to\Psires^{m}$ and $S^m_{\Psi}\to S^m_{\Psi{\rm eff}}$.

For our purpose, we rewrite the $m$-mode field equation (\ref{Boxm}) as two coupled first-order equations in the (complex-valued) variables $\Psi^m$ and $\Pi^m \equiv \partial \Psi^m / \partial t$. Explicitly,
\begin{eqnarray}
\Psi^m_t &=& \Pi^m,  \nonumber \\
\Pi^m_t   &=& - \frac{4 i a m M r}{\Sigma^2} \Pi^m  + \frac{(r^2 + a^2)^2}{\Sigma^2}  \Psi^m_{r_\ast r_\ast} + \left[\frac{2iamr(r^2+a^2) - 2 a^2 \Delta}{r \Sigma^2} \right] \Psi^m_{r_\ast} \nonumber  \\ \nonumber
 && + \frac{\Delta}{\Sigma^2} \Psi^m_{\theta \theta} + \frac{\Delta \cot \theta}{\Sigma^2 }\Psi^m_{\theta} - \frac{\Delta}{\Sigma^2} \left[\frac{m^2}{\sin^2\theta} 
+\frac{2M}{r}\left(1-\frac{a^2}{Mr}\right) 
  + \frac{2 i a m}{r} \right] \Psi^m + S^m_{\Psi} ,  \label{firstorder-1} 
\end{eqnarray}
where $\Psi^m_t$, $\Psi^m_{r_\ast}$, $\Psi^m_{\theta}$ etc.~denote partial derivatives. This form is suitable for the application of the {\it method of lines}, which we adopt here. (The method is reviewed, e.g., in Ref.~\cite{Schiesser} and in Sec.~4.2 of Ref.~\cite{Rinne}.) 

\begin{figure}
 \includegraphics[width=8cm]{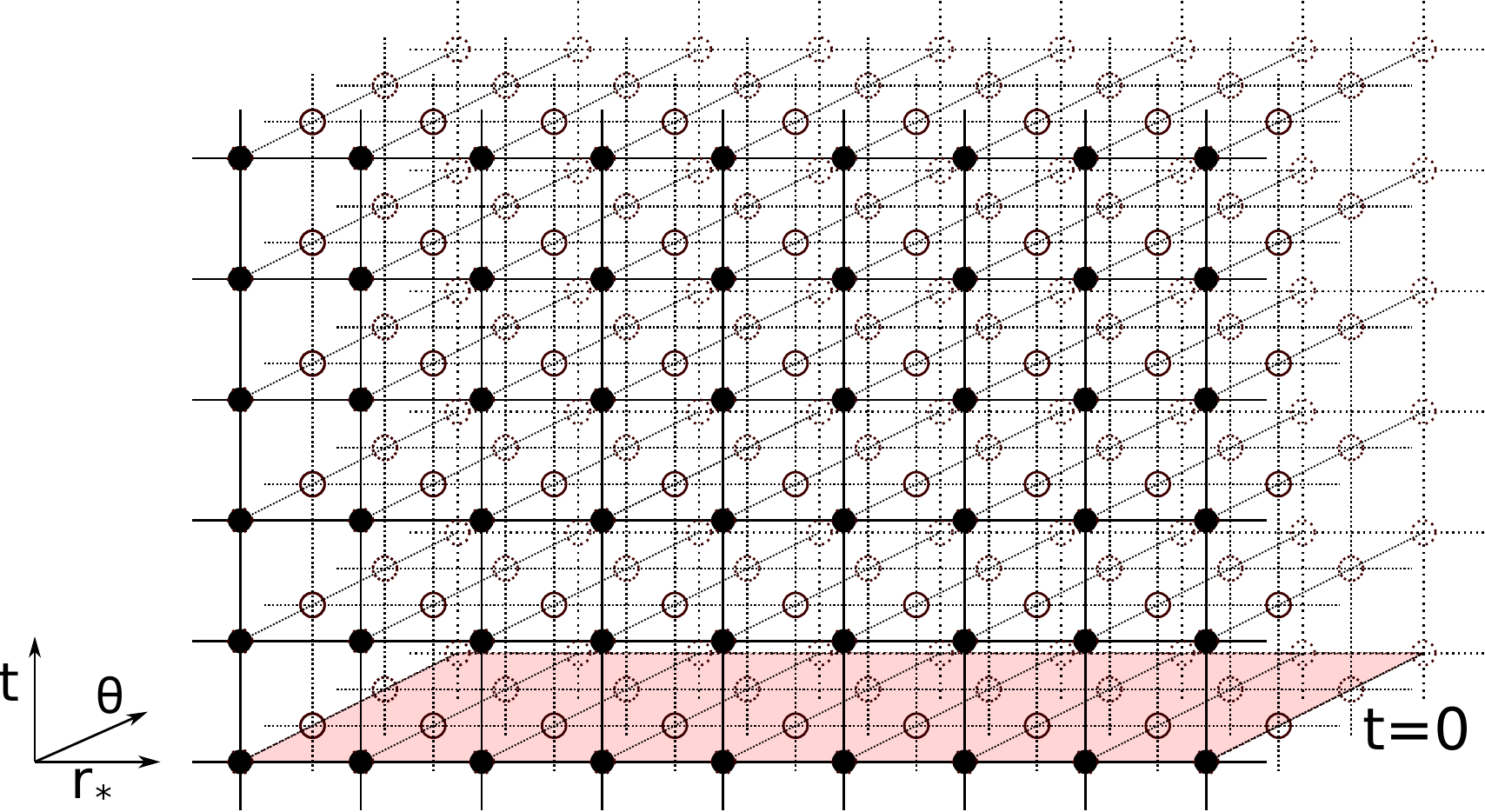}
 \includegraphics[width=6cm]{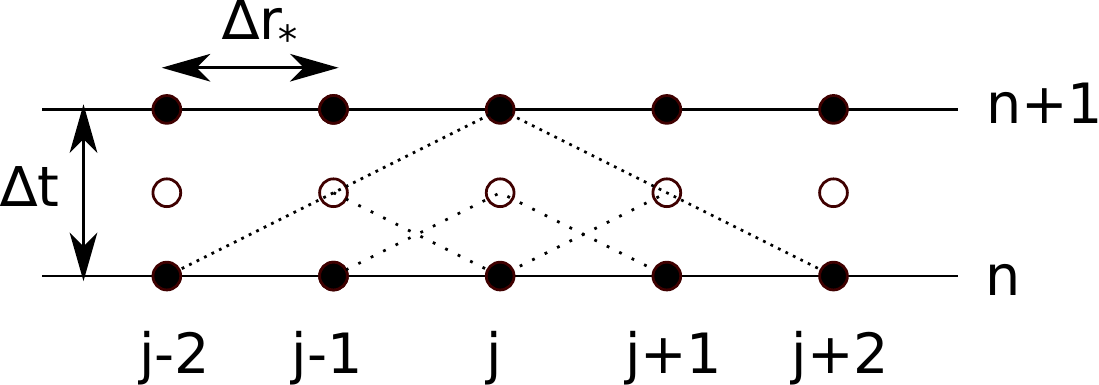}
 \caption{Grid for finite difference method. The left plot shows the 2+1D grid with linear spacings ($\Delt$, $\Delr$, $\Delq$). The evolution proceeds from initial data on the surface $t=0$ (shaded). The right plot illustrates a grid cell for the method-of-lines finite difference method (with the $\theta$ direction suppressed for clarity). To obtain the value at point $(j, n+1)$, we first use the values at the 5 grid points $(j-2,n), \ldots, (j+2,n)$ to obtain estimates on an intermediate time slice (unfilled circles); cf.\  Eq.~(\ref{eq:mol}). The domain of dependence of the grid point at $(j,n+1)$ is shown with (densely) dotted lines. }
 \label{fig:grid}
\end{figure}

Let $\Psi_{njk} \equiv \Psi^m(t_n, r_{\ast j}, \theta_k)$ denote the field value at the grid point corresponding to the coordinate values
\beq
t_n = n \Delt, \quad \quad r_{\ast j} = r_{\ast0} + j \Delr , \quad \text{and} \quad \theta_k = k \Delq .
\eeq
To write Eqs.\ (\ref{firstorder-1}) in an FD form, we replace $\Psi^m   \rightarrow \Psi_{njk}$, 
and replace all spatial derivatives (in $r_\ast$ and $\theta$ coordinates) with finite-difference approximations of second-order accuracy:
\begin{eqnarray}
\Psi^m_{r_\ast} &\rightarrow& \frac{ \Psi_{n(j+1)k} - \Psi_{n(j-1)k}  }{2 (\Delr)} , \\
\Psi^m_{r_\ast r_\ast} &\rightarrow& \frac{ \Psi_{n(j+1)k} -  2 \Psi_{njk} + \Psi_{n(j-1)k}  }{(\Delr)^2} , \\
\Psi^m_{\theta} &\rightarrow& \frac{\Psi_{nj(k+1)} - \Psi_{nj(k-1)}}{2 (\Delq)}, \\
\Psi^m_{\theta\theta} &\rightarrow& \frac{ \Psi_{nj(k+1)} -  2 \Psi_{njk} + \Psi_{nj(k-1)}  }{(\Delta\theta)^2} . 
\end{eqnarray}
To evolve the equations in time (i.e., to step forward to obtain $\{ \Psi_{(n+1)jk}, \Pi_{(n+1)jk} \}$ from $\{ \Psi_{njk}$, $\Pi_{njk} \}$) we applied a fourth-order Runge--Kutta step. To illustrate the method in general, consider a coupled set of first-order equations of the form
\beq
\partial_t \bu = \bF(t, \bu) ,
\eeq
where $\bu$ and $\bF$ are vectors. We denote the set of unknowns $\bu$ at time $t_n$ as $\bu_n$ (in our case, $\bu_n = \left\{ \Psi_{njk} , \Pi_{njk} \right\}$ for all values of $j,k$). To obtain an approximation $\bu_{n+1}$ at time $t_{n+1} = t_n + \Delt$ we apply
\beq
\bu_{n+1} = \bu_n + \Delt \left( b_1 \bk_1 + b_2 \bk_2 + b_3 \bk_3 + b_4 \bk_4 \right),
\label{eq:mol}
\eeq
which combines information from a sequence of intermediate estimates,
\begin{eqnarray}
\bk_1 &=& \bF\left(t_n, \bu_n\right), \\
\bk_2 &=& \bF\left(t_n + c_2 \Delt, \bu_n + \Delt \, a_{21}  \bk_1\right), \\
\bk_3 &=& \bF\left(t_n + c_3 \Delt, \bu_n + \Delt \, (a_{31} \bk_1 + a_{32} \bk_2) \right), \\
\bk_4 &=& \bF\left(t_n + c_4 \Delt, \bu_n + \Delt \, (a_{41} \bk_1 + a_{42} \bk_2 + a_{43} \bk_3) \right).
\end{eqnarray}
We used a standard version of the 4th-order Runge--Kutta method, in which the coefficients are $a_{21} = a_{32} = \tfrac{1}{2}$, $a_{43} = 1$, $a_{31} = a_{41} = a_{42} = 0$, $b_1 = b_4 = \tfrac{1}{6}$, $b_2 = b_3 = \tfrac{1}{3}$, $c_2 = c_3 = \tfrac{1}{2}$ and $c_4 = 1$. 
Note the function $\bF(\cdot)$ couples between $\left\{ \Psi_{njk} , \Pi_{njk} \right\}$ and 
$\left\{ \Psi_{n(j\pm 1)(k\pm 1)} , \Pi_{n(j\pm 1)(k\pm 1)} \right\}$; see the right panel of
Fig.~\ref{fig:grid}, which illustrates the method.

The above scheme is second-order convergent in the spatial resolution and fourth-order convergent in the time resolution. We opted for a 4th-order convergence in time scheme because the method of lines with a 2nd-order Runge--Kutta time step is known to be unconditionally unstable \cite{Rinne}, and (through experimentation) we found that a 3rd-order scheme requires an impractically small time step to remain stable. We comment more on stability in Sec.\ \ref{subsec:stability} below.

\subsection{Initial and boundary conditions\label{subsec:ic-bc}}
We specify a ``zero'' initial condition on the initial surface at $t=0$, i.e., $\Psi^m_{0jk} = 0 =\Pi^m_{0jk}$ for all $j,k$. Of course, this is not a solution of the sourced field equations; however (as discussed in Paper I and Ref.~\cite{Jaramillo:2011gu}) initial ``junk radiation'' is found to radiate away at late times. We monitor the level of this transient radiation on the worldline (by measuring the deviation from stationarity), and discard the early part of the evolution, where the magnitude of this radiation is high. The residual ``contamination'' (due to imperfect initial conditions) diminishes with time, but may remain a significant source of numerical error. We discuss this further in Sec.\ \ref{m-modes}.


The physical boundary conditions to be applied at the poles ($\theta = 0$ and $\theta = \pi$) are \cite{Barack:Golbourn:2007}, 
\beq
\Psi_{\theta}^{m=0}(\theta \in \{ 0, \pi \}) = 0,  \quad \quad \Psi^{m \neq 0}(\theta \in \{ 0, \pi \}) = 0 . \label{polar-bc1} 
\eeq
To implement these conditions, we simply set $\Psi^m = \Pi^m= 0$ at the poles for $m\neq0$, and for $m=0$ we extrapolate to obtain
\begin{eqnarray}
\Psi^{m=0}(\theta = 0) &=& \frac{1}{3} \left[ 4 \Psi^{m=0}(\theta=\Delq) - \Psi^{m=0}(\theta = 2 \Delq) \right] +O\left( (\Delq)^4 \right),  \label{polar-bc2} \\
\Psi^{m=0}(\theta = \pi) &=& \frac{1}{3} \left[  4 \Psi^{m=0}(\theta = \pi - \Delq) - \Psi^{m=0}(\theta = \pi - 2\Delq) \right] + O \left( (\Delq)^4 \right) ,   \label{polar-bc3}
\end{eqnarray}
and likewise for $\Pi^m$. Note that the error term is $O((\Delq)^4)$, rather than $O((\Delq)^3)$, because $\Psi^m$ is an even function of $\theta$ at $\theta=0$.

Conditions are also required at the boundaries of the grid in $r_\ast$, denoted $r_{\ast\text{min}}$ and $r_{\ast\text{max}}$. A correct but technically demanding approach is to impose ``absorbing'' boundary conditions, i.e., purely ingoing radiation at the horizon and purely outgoing radiation at spatial infinity. We found it difficult, however, to control sufficiently well the level of initial spurious radiation bouncing off the spatial boundaries and contaminating the late-time solution. Instead, we use a simple ``zero-derivative'' condition, $\Psi_{r_\ast}^m = \Pi_{r_\ast}^m = 0$ at the boundaries $r_{\ast\text{min}}$ and $r_{\ast\text{max}}$, and merely ensure that the grid is of sufficient span in $r_\ast$ that the initial burst of spurious radiation does not have time to reflect from the boundary and re-encounter the worldline. This approach is less computationally economical but has the advantage of simplicity.

\subsection{Numerical stability of method\label{subsec:stability}}


Our FD strategy here differs from that of Paper I, where we used a semi-characteristic grid in coordinates $u = t - r_\ast$, $v = t + r_\ast$ and $\theta$ (in Schwarzschild). Let us refer to this as the ``$uv\theta$'' method. It is less natural to apply the $uv\theta$ method in the Kerr case, where the presence of a first time derivative in the field equation complicates the scheme considerably. In fact, we were not able to obtain sufficiently stable evolutions based on the $uv\theta$ grid and thus abandoned the approach in favour of the method of lines. 

In the Schwarzschild case we found that the $uv\theta$ method was subject to a numerical instability (manifested in an exponential growth of high-frequency noise) if the angular spacing $\Delq$ was set to be too small. More specifically, the method was found to be unstable if \cite{Dolan:Barack:2011}
\beq
\frac{\Delq}{h}  \le  \frac{1}{2} \text{max} \left( r^{-2} \Delta^{1/2} \right) \sqrt{1 + m^2 / 4} ,
\eeq
where $h = \triangle u = \triangle v$ is the characteristic grid spacing, $m$ is the azimuthal mode number, and in the Schwarzschild case $\Delta=r^2-2Mr$. The instability was seen to arise first near the poles, and we showed that it could be mitigated by moving the polar boundaries inward (see Sec.~III.C.5 of Paper I). 
The version of the method of lines described in Sec.~\ref{subsec:FDmethod} suffers from a similar angular instability. Empirically, we find that an instability arises if 
\beq \label{instab-RK}
\frac{\Delq}{\Delr}  \lesssim    \text{max} \left( r^{-2} \Delta^{1/2} \right) \sqrt{1 + m^2 / 4}  ,  
\eeq
for $\Delt = \Delr$. Again, moving the polar boundaries inward (particularly for the large-$m$ modes) mitigates this problem, as described in Paper I. 

Explicit FD methods may also suffer from ``radial'' instabilities if the ratio $\nu \equiv \Delt / \Delr$ is taken too large. We chose to implement the method of lines with a 4th-order Runge--Kutta time step primarily for its known good stability properties \cite{Rinne}. We were able to use the ratio $\nu = 1$ without problems. Below we comment on some experiments with other methods.


\subsection{Alternative finite-difference methods}
We have experimented with several alternative FD methods for evolving the scalar field on the Kerr spacetime in 2+1D, some of which have already been used in the literature. We report here briefly on our findings.   

Krivan \emph{et al.} \cite{KLPA97, KLP96} (see also Sundararajan \emph{et al.} \cite{SKH07}) rewrote Teukolsky's master scalar-field equations in 2+1D as a coupled set of first-order equations in variables $\Psi^m$ and $\tilde{\Pi}^m$, where
\beq
\tilde{\Pi}^m = \frac{\partial \Psi^m}{\partial t} + \frac{r^2+a^2}{\Sigma}  \frac{\partial \Psi^m}{\partial r_\ast} .
\eeq
They applied a two-step Lax-Wendroff method \cite{NumericalRecipes} to obtain a second-order-accurate algorithm to evolve forward in time. In our experiments with this method, we found that an angular instability (Sec.~\ref{subsec:stability}) arises if $\Delq / \Delr \lesssim \pi \max(r^{-2} \Delta^{1/2}) \sqrt{1 + m^2/8}$ (for $a = 0$ and $\Delt = \Delr$), and becomes rapidly more severe with increasing $|a|$. Note that at large $m$ this necessitates fixing the ratio $\Delq / \Delr$ at larger values than in our method of lines [compare with Eq.\ (\ref{instab-RK})]. To achieve a given resolution in the angular direction therefore requires smaller values of $\Delr$ (and $\Delt$), which increases the computational cost.

In addition, we tried a version of Krivan \emph{et al.}'s approach using a leapfrog method \cite{NumericalRecipes} in place of the two-step Lax-Wendroff. With this approach, we found that, at least in the Schwarzschild limit, the angular stability condition (\ref{instab-RK}) applies, just as in our method of lines. However, we also found that in the Kerr case ($a \neq 0$), relatively small values of $\nu$ were required to obtain stable runs.

We have also experimented with replacing the 4th-order Runge--Kutta step in the method of lines with either (i) a 3rd-order Runge--Kutta step, or (ii) an iterative Crank--Nicholson step \cite{NumericalRecipes}. In either case, the stability properties were found to be more restrictive than with the 4th-order Runge--Kutta method.

Our numerical experiments have been restricted to coordinate grids in $t, r_\ast, \theta$. Other authors have highlighted benefits of working with alternative coordinate foliations of black hole spacetimes \cite{PeterDiener,Zenginoglu:2008}. For example, Zenginoglu \cite{Zenginoglu:2011} has discussed how, by making a suitable replacement of the time coordinate, $t = \hat{t} + \alp(r)$, one may use the method of lines on constant-$\hat{t}$ hypersurfaces that are asymptotically null and intersect the event horizon and future null infinity. Since the fields $\Psi^m$ tend towards constants on such surfaces, one may use a compactification step. R\'acz and T\'oth \cite{Racz:Toth:2011} recently presented their own implementation of a similar idea for the Kerr spacetime. 

\subsection{Simulations, data extraction and numerical error}

For a given mode $m$ we evolve the initial data (Sec.~\ref{subsec:ic-bc}) according to the finite difference scheme (Sec.~\ref{subsec:FDmethod}) with boundary conditions (\ref{polar-bc1})--(\ref{polar-bc3}) on a 2+1D grid (Fig.~\ref{fig:grid}) in the range $0 \le t \le \tmax$. The numerical solution depends on ``physical'' parameters, $r_0$, $a$ and $m$, and a set of ``numerical'' parameters, $\{\text{num.}\} = \{\Delt, \Delr, \Delq, \Gamma_{r_\ast}, \Gamma_\theta, \tmax, \ldots \}$. We refer to a simulation for a given $m$, $r_0$, $a$ with a unique set of $\{\text{num.}\}$ as a ``run''.  

In Sec.~IIID of Paper I, we described the procedure for computing the SF from a set of runs; let us recap the steps here. First, we ensure that $\tmax$ is sufficiently large that the field inside the worldtube has settled into a stationary state (see Sec.~\ref{sec:results} below). Next, we read off the following estimates for the modal contributions to the $r$ and $\phi$ components of the SF (for $m\geq 0$):
\begin{eqnarray}
F_r^m(t) &=& r_0^{-1} \left[ \frac{r_0^2 + a^2}{\Delta_0} \partial_{r*} \Psireshat^m - r_0^{-1} \Psireshat^m \right] (t, r_0, \pi/2, \Omega t),  \label{num-Fr}  \\
F_\phi^m(t) &=& - 2 m r_0^{-1} \text{Im} \left[ \Psires^m(t,r_0,\pi/2) e^{i m \Omega t} \right],   \label{num-Fphi}
\end{eqnarray}
where $\Delta_0\equiv \Delta(r=r_0)$, and where we have defined the (real) ``total $m$-mode contribution'' via
\beq
\Psireshat^m(t,r,\theta,\varphi) = \begin{cases} 2 \text{Re} \left[ \Psires^m(t,r,\theta)e^{i m \varphi}  \right] , \quad & m > 0, \\ \Psires^m(t,r,\theta), & m = 0. \end{cases}
\label{Psirestil}
\eeq
Here all quantities are evaluated at grid points on the worldline at late times $t < \tmax$, and the derivative with respect to $r_\ast$ is found via central differencing. In principle, the SF components are then found via the mode sums
\beq
F^{\text{self}}_r = \sum_{m = 0}^\infty F_r^m , \quad \quad F^{\text{self}}_\phi = \sum_{m = 0}^\infty F_\phi^m  ,
\eeq
along with $F^{\text{self}}_t = - \Omega F^{\text{self}}_\phi$ and, by symmetry, $F^{\text{self}}_\theta = 0$. In practice, the modal values (\ref{num-Fr})--(\ref{Psirestil}) extracted from a particular run depend somewhat upon the set of numerical parameters $\{\text{num.}\}$. In other words, the modal values contain numerical error, which we define as the difference between the numerical solution and the (unknown) exact solution.

Several key sources of numerical error were discussed in Sec.~IIIE of Paper I. To recap, we have: (i) discretization error, associated with  the use of a finite grid spacing $\Delt, \Delr, \Delq$; (ii) worldtube error, which varies depending on the dimensions of the worldtube, $\Gamma_{r_\ast}, \Gamma_{\theta}$; (iii) relaxation-time error, associated with nonstationary residual spurious radiation; and (iv) source cancellation error, arising from small inaccuracies in the numerical evaluation of the effective source near the worldline due to delicate mutual cancellation of terms.  There arise additional errors in computing the mode sums: (v) $m$-mode summation error, due to imposing a finite large-$m$ cut-off $\mmax$, and fitting the remainder with an approximate model; and (vi) mode cancellation error, which can arise from delicate mutual cancellation of individual $m$-mode contributions leading to a high relative error in the mode sum. 
 
In Paper I we described in some detail how to monitor, control and (to an extent) mitigate these sources of error. Our treatment is very similar in the Kerr case. In Sec.~\ref{sec:results} we will illustrate the effect of the various sources of error in our particular Kerr implementation.

\subsection{Computational resource}
Numerical error can be reduced simply by applying greater computing resources to the problem. For example, errors (i) and (ii) diminish as we increase the resolution (i.e., decrease $\Delt, \Delr, \Delq$), and error (iii) diminishes if we run the simulation for longer (i.e., increase $\tmax$). However, the runtime scales as $\tmax^2 / ( \Delt \Delr \Delq)$ and so in practice we soon encounter a practical limit. 

Luckily, our numerical algorithm lends itself easily to parallelization, with each run (for given $m$, $a$ $r_0$, $\{\text{num.}\}$) being assigned to an independent thread. To facilitate this parallel computation we make use of the {\sc Iridis} 3 HPC resource. The SF is computed by post-processing the results of multiple runs, as described in the next section. Typically, to compute $F^{\text{self}}_r$ and $F^{\text{self}}_\phi$ for a given $r_0$ and $a$, we computed the modes $m = 0, \dots, 19$ up to $\tmax = 300M$ at four different resolutions, $\Delr / M = 1/8, 1/16, 1/24, 1/32$ [with fixed ratios $\Delt / \Delr = 1$ and $\Delq / \Delr = \pi / (6M)$]. Each of the 80 runs is assigned to a separate node. Some additional resource is devoted to the $m=0$ mode, to ensure a fuller late-time relaxation of this slowest-relaxing mode (see below); typically, we run this mode up to $\tmax = 500M$. 

The longest runtime for one complete simulation in this work (i.e., one computation of the SF with given $r_0$, $a$)  was $\sim 24$ hours.


\section{Results and Analysis\label{sec:results}}
In this section we present a selection of results from our numerical simulations, and we discuss the
challenge of minimizing numerical errors. We compare the results with those obtained in \cite{Warburton:Barack:2010}, and find good agreement to within the estimated numerical error. 

\subsection{$m$-modes}\label{m-modes}
Let us start by considering a typical run, i.e., a single simulation with given $r_0$, $a$, and $m$, and with a particular set of numerical parameters. We can visualize the run by extracting data along three slices: (i) $t = \tmax$, $\theta  = \pi/2$, i.e., in the equatorial plane, (ii) $t = \tmax, r_\ast = r_{\ast0}$, i.e., from pole to pole, crossing the worldline, and (iii) $r_\ast = r_{\ast0}$, $\theta = \pi/2$, i.e., along the worldline.

\begin{figure}
  \begin{center}
  \includegraphics[width=8.1cm]{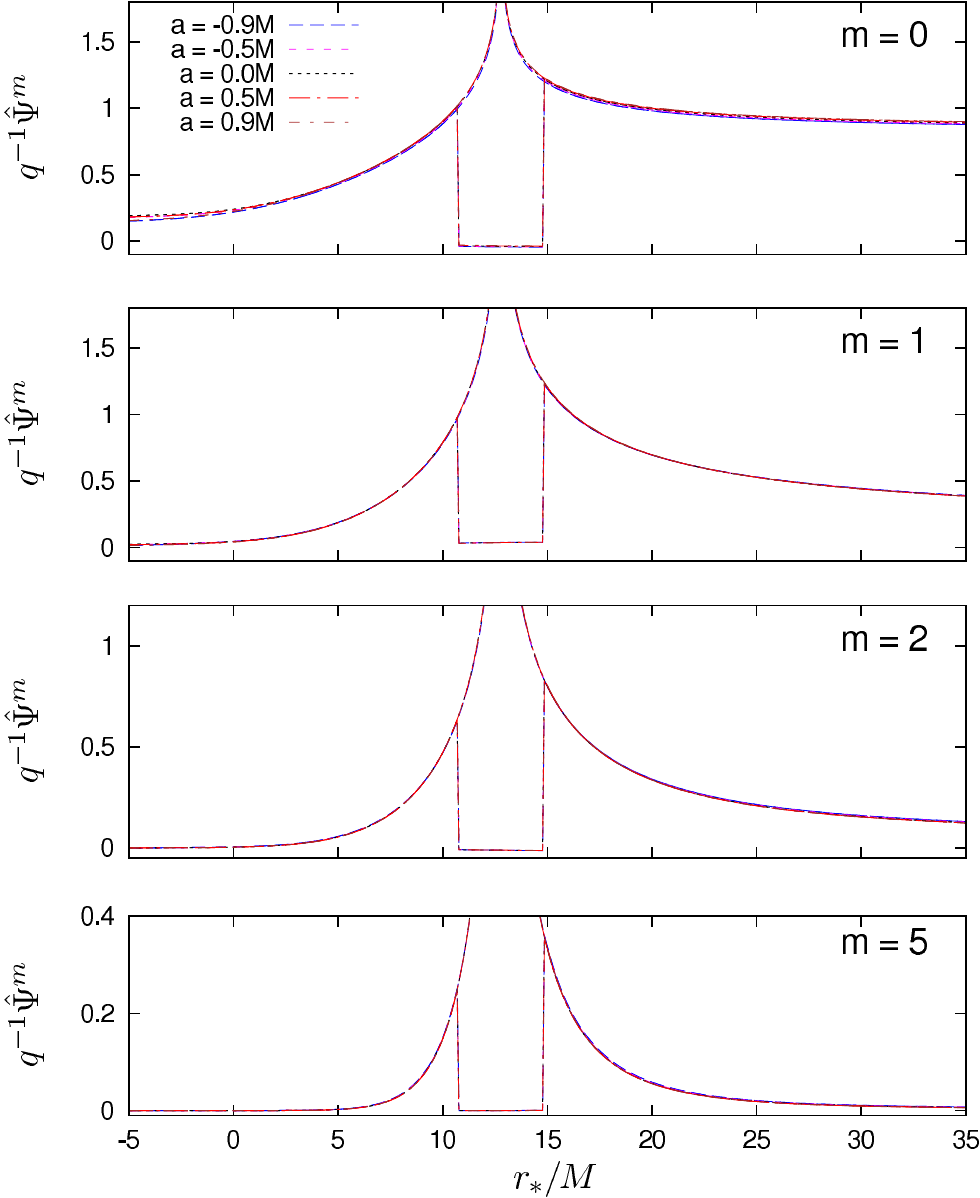}
  \includegraphics[width=8.1cm]{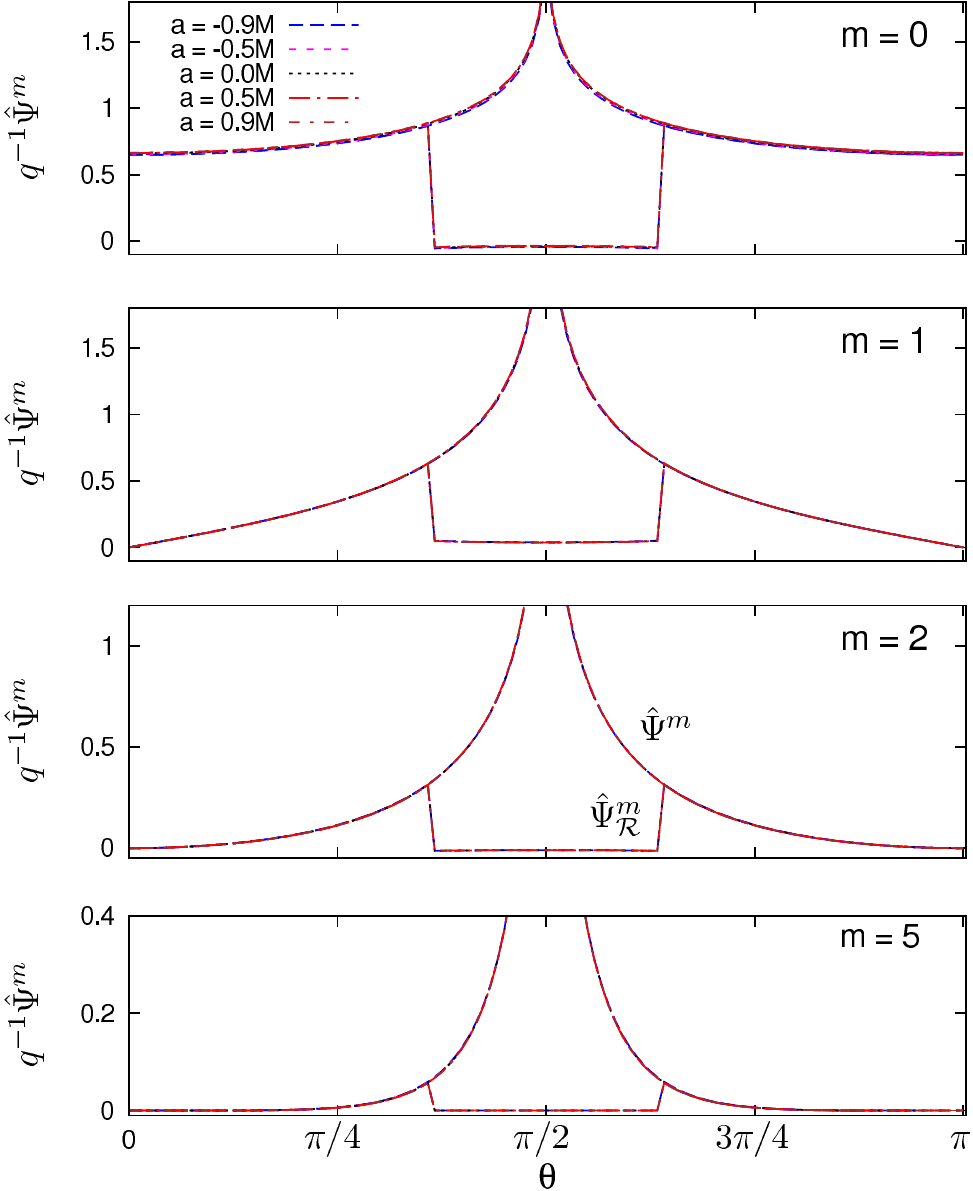}
 \end{center}
\caption{Field modes ($m=0,1,2,5$) on a constant time slice (at $t = \tmax$) for circular orbits at $r_0 = 10M$, for a range of Kerr parameters ($a/M = -0.9, -0.5, 0, 0.5, 0.9$). The left plots show field modes at fixed $\theta = \pi/2$ and the right plots show field modes at fixed $r=r_0$. Inside the worldtube we show both the residual field $\Psirestil^m$ (forming the `trough') and the full retarded field $\Psirettil^m$, which is divergent on the worldline. We note that the rotation rate $a/M$ has only a subtle effect on the field profile.}
 \label{fig:profiles}
\end{figure}

Fig.~\ref{fig:profiles} shows typical $m$-mode contributions to the field along the constant-$t$ slices (i) and (ii), for an orbit at $r_0 = 10M$ and a range of Kerr parameters $a$. The worldtube is visible as a central `trough'; inside the tube, we show both the residual $\Psirestil^m$ and the full field $\Psirettil^m$ (which diverges logarithmically as $r \rightarrow r_0$, $\theta \rightarrow \pi/2$). These plots are similar to those for the Schwarzschild implementation (see Fig.~4 in Paper I). We note that the effect of black hole rotation upon the field mode profiles is quite subtle, although it has a more profound effect on the SF.

\begin{figure}
  \begin{center}
  \includegraphics[width=16cm]{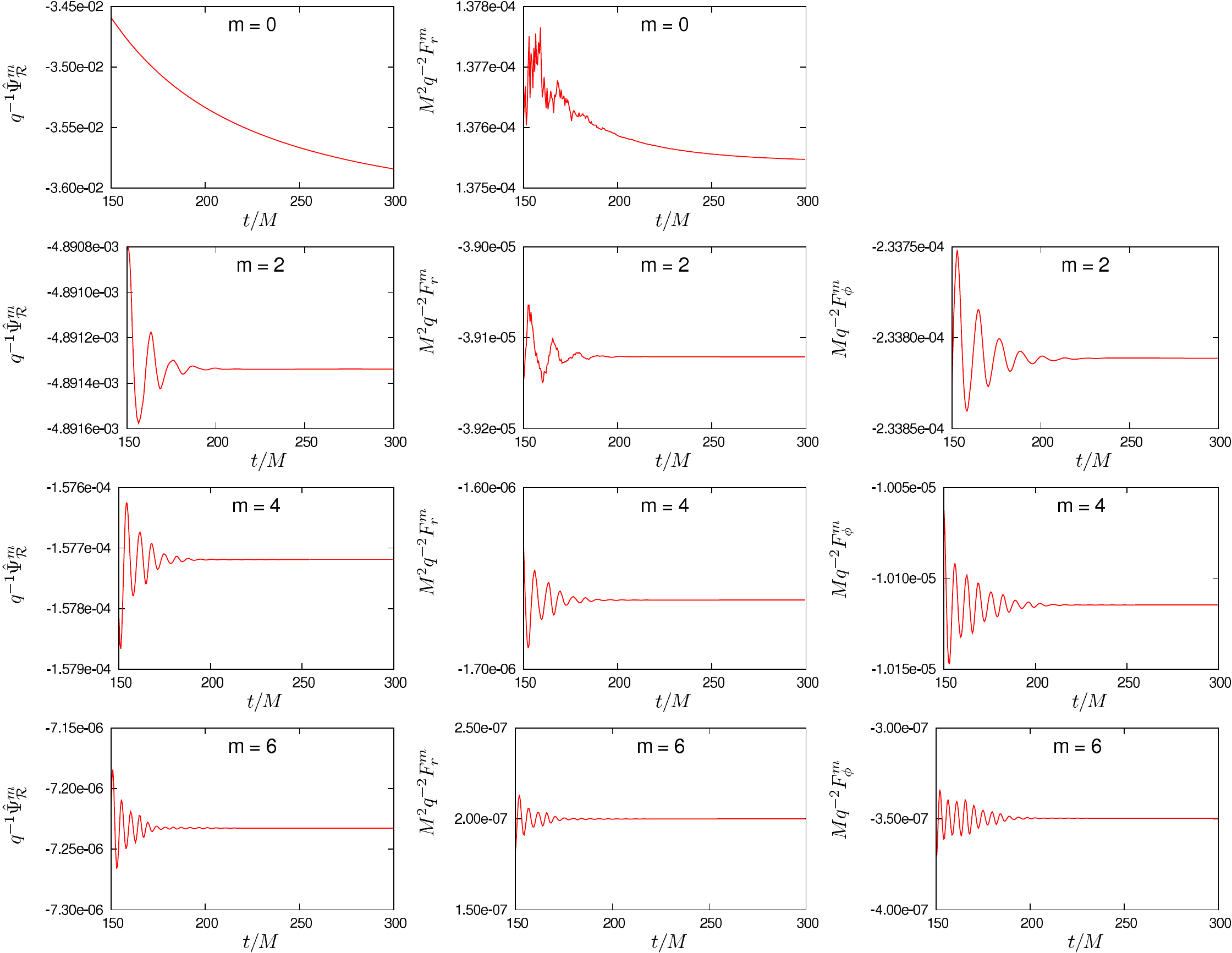}
 \end{center}
\caption{Typical time series data for the modes of the field $\Psirestil$ (left column), the radial SF $F_r$ (center column) and the angular SF $F_\phi$ (right column) for a range of $m$ modes (top to bottom: $m = 0, 2, 4, 6$). These data are taken from simulations with radius $r_0 = 10M$ and Kerr parameter $a=0.5M$, with numerical parameters $\Delt = \Delr = M / 32$, $\Delq = \pi  / 192$ and $\Gamma_{r_\ast} = 2.5M$, $\Gamma_\theta = \pi / 4$. Note that $F_\phi^{m=0}$ is identically zero.}
 \label{fig:time-series}
\end{figure}

Figure \ref{fig:time-series} shows plots of $\Psirestil^m$, $F_r^m$ and $F_\phi^m$ as functions of $t$ on the worldline [i.e., on slice (iii)], for runs with $r_0 = 10M$, $a = 0.5M$ and modes $m = 0$, $2$, $4$ and $6$. After an initial burst of junk radiation (due to imperfect initial conditions, Sec.~\ref{subsec:ic-bc}), the modal quantities settle towards steady-state values. Visible in the figures are two types of transients: Initially, there are regular high-frequency oscillations (for $m \neq 0$) which may be identified as quasi-normal ringing (indeed,the ringing frequency is proportional to $m$ as expected, and the exponential decay rate of the ringing seems roughly independent of $m$, also as expected). At later times the modes exhibit a second type of transient behavior: a power-law decay with an $m$-dependent exponent.
In Paper I (Sec.~IVA5 with, e.g., Fig.~10) we explored this power-low behavior in some detail, and demonstrated that, by fitting the decay of the field with an asymptotic model, we can extrapolate to $t \rightarrow \infty$ to extract a steady-state value. We implement the same method here.

\subsection{Convergence with grid resolution\label{subsec:convergence}}
For simplicity, let us fix the ratios $\Delr / \Delt$ and $\Delq / \Delt$ and vary only $\Delt$. We will refer to some quantity $X(\Delt)$ as being ``$k$th-order convergent'' if $X(\Delt)$ converges to some finite value as $\Delt \rightarrow 0$ in the following way:
\beq
X(\Delt) = X(\Delt = 0) + O(\Delt^k) . 
\eeq
To test the rate of convergence $k$ of some quantity $X(\Delt)$ [e.g., the residual field modes $\Psirestil^m(\Delt)$ or the SF modes $F_{r/\phi}^m(\Delt)$], we may take the ratios of the differences between the results of runs with different resolutions. A standard test is to examine the quantity
\beq
\chi(\Delt)  = \frac{X(2 \Delt) - X(\Delt)}{X(\Delt) - X(\Delt/2)} .   \label{chi-def}
\eeq
If the method is $k$th-order convergent then $\chi(\Delt)$ will approach $2^k$ in the limit $\Delt \rightarrow 0$.

In our implementation of the method of lines we use 2nd-order-accurate finite differencing on the spatial slices. Hence, we expect to find that $\Psi^m$ is 2nd-order convergent. In addition, we expect the SF quantities $F_r^m$ and $F_\phi^m$ also to be 2nd-order convergent. 
In Table~\ref{tbl-conv-test} we give a few sample values of $\chi$ from our simulations, for a median resolution of $\Delt = M/16$, and ratios $\Delr / \Delt = 1$ and $\Delq / \Delt = \pi/(6M)$. We find that these values fall in the range $3.7 < \chi < 4.3$ at this resolution, suggesting a second-order convergence of our FD scheme, as expected. We verified that, as the resolution is increased (i.e., $\Delt, \Delr, \Delq$ are decreased) the results of the convergence test improve (i.e., $\chi$ becomes closer to $4$). We find that, in general, the convergence deteriorates somewhat at large values of $m \gtrsim 15$. 

\begin{table}
\begin{tabular}{l l | c   c }
\hline\hline
$a/M$ \hspace{2mm} & $m$ \hspace{2mm} & $\chi$ (for $F_r^m$) \hspace{2mm} & $\chi$ (for $F_\phi^m$) \\
\hline
$0$ & $ 3$ &  4.17 &   3.89 \\
&$ 6$ & 3.96 &  3.94 \\
&$ 9$ & 3.82 & 3.99 \\
\hline
$0.5$ & $ 3$ & 4.17 & 3.94 \\
&$ 6$ & 3.97 & 3.85 \\
&$ 9$ & 3.82 & 3.74 \\
\hline
$0.9$ & $ 3$ & 4.19 & 4.00 \\
&$ 6$ & 4.00 & 3.83 \\
&$ 9$ & 3.77 & 4.14 \\
\hline \hline
\end{tabular}
\caption{Sample convergence tests for $r_0 = 6M$, $a = 0.5M$, showing numerical values for $\chi$ as defined in Eq.~(\ref{chi-def}) for the median grid spacings $\Delr = \Delt = M/16$ and $\Delq = \pi/96$, with $\tmax = 300M$. }
\label{tbl-conv-test}
\end{table}

\begin{figure}
  \begin{center}
  \includegraphics[width=0.49\columnwidth]{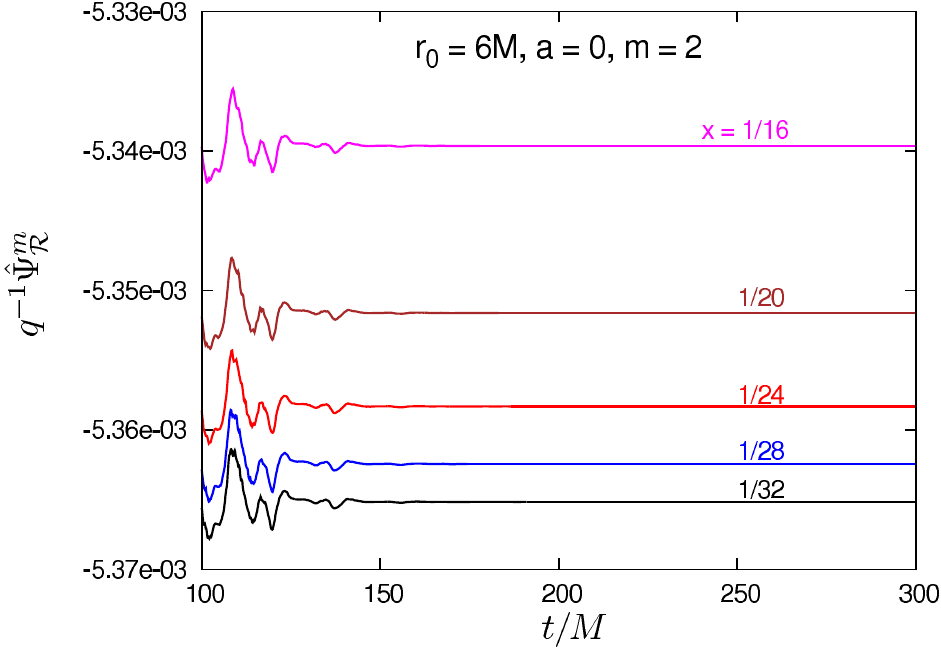}
  \includegraphics[width=0.49\columnwidth]{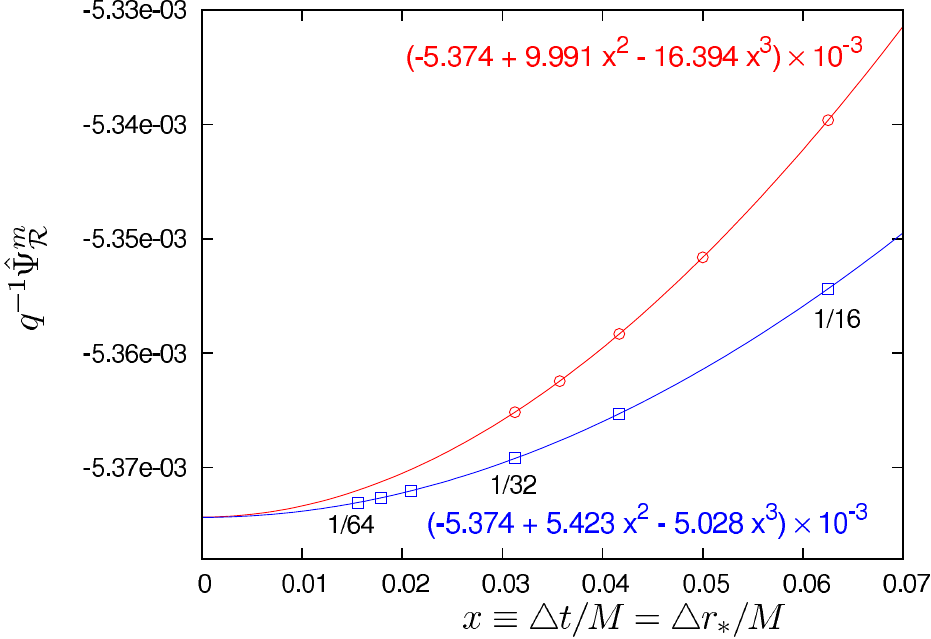}
 \end{center}
\caption{Convergence of the FD scheme and Richardson extrapolation. The left plot shows the $m=2$ mode of the residual field on the worldline, $\Psirestil^m$, for $a=0$ and a circular orbit at $r_0 = 6M$, as a function of time. The lines show the results from simulations with various grid resolutions, $x = 1/16$, $1/20$, $1/24$, $1/28$ and $1/32$, where $x \equiv \Delt / M = \Delr / M = 6 \Delq / \pi$. The right plot shows the values at $\tmax = 300M$ as a function of $x$. The red circles on the upper line show the values taken from the simulations in the left plot. The blue squares on the lower line show the values from the $uv\theta$ method (which only works for $a=0$). The two data sets have been fitted with cubic polynomial models, $\Psirestil^m=c_0 + c_2 x^2+ c_3 x^3$ (red and blue lines). Reassuringly, the values $c_0$ extrapolated from the two data sets (which correspond to the zero grid spacing limit) are found to be in very good agreement. }
 \label{fig:convergence}
\end{figure}

Once we have confirmed that the code is 2nd-order convergent, we may apply a Richardson extrapolation (``Richardson's deferred approach to the limit'' \cite{NumericalRecipes}) to reduce the discretization error. Figure \ref{fig:convergence} illustrates the idea. The left plot shows numerical results for the field on the worldline, as a function of simulation time $t$, for various resolutions $1/16 \le x\equiv \Delt / M \le 1/32$ (and $a=0$). The right plot shows the values at $t = 300M$ as a function of resolution, and compares with a similar data set obtained using the $uv\theta$ method (which is only valid in the $a=0$ case). By fitting the data set with a simple polynomial model, i.e.
\beq
\Psirestil^m(x) = \Psirestil^m(x=0) + c_2 x^2 + c_3 x^3 ,
\eeq
we may extrapolate to infinite resolution ($x = 0$). Figure \ref{fig:convergence} shows that the extrapolated values from the two data sets are in excellent agreement. The difference between the extrapolated values from two data sets (for example, runs with different worldtube dimensions) provides a rough estimate of the error in the fitting procedure.

\subsection{Mode sums\label{subsec:modesums}}
\begin{figure}
  \begin{center}
  \includegraphics[width=8cm]{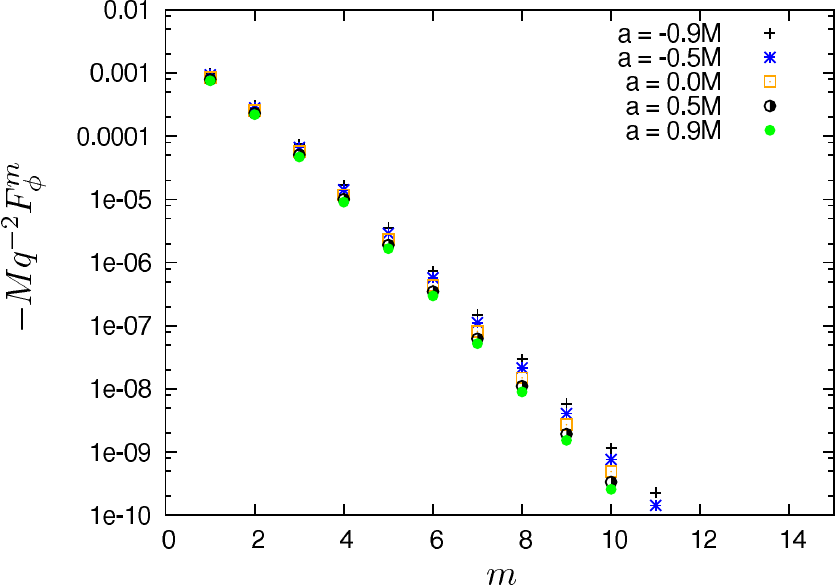}
  \hspace{0.2cm}
  \includegraphics[width=8cm]{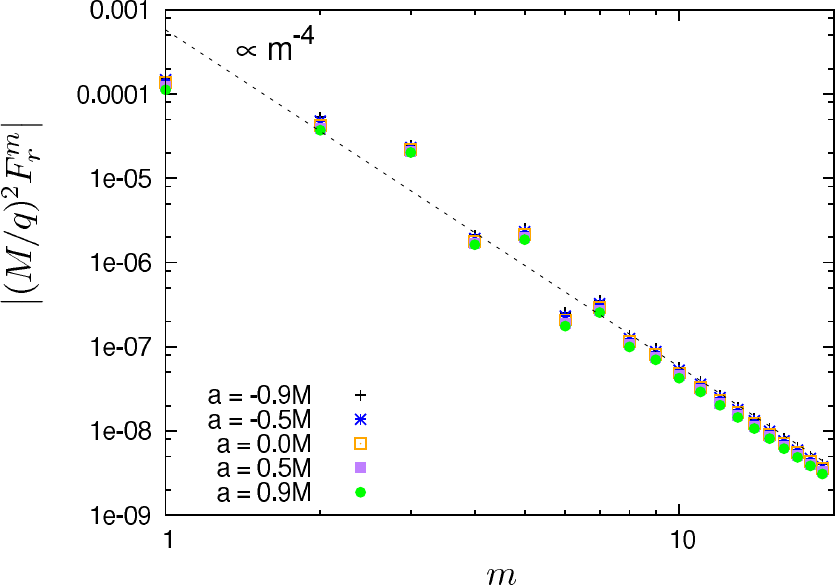}
  \includegraphics[width=8.1cm]{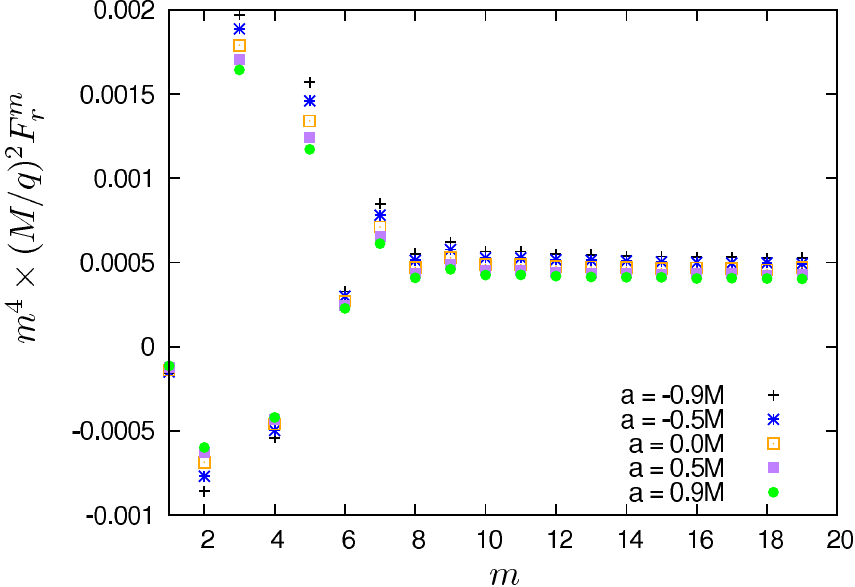}
 \end{center}
\caption{Convergence of modal contributions as a function of $m$, for a 4th-order puncture implementation ($r_0 = 10M, \tmax = 300M$). The upper left plot shows (minus) the $m$ modes of the angular (dissipative) component $F_\phi$, on a semilog scale. We observe exponential convergence with $m$. The upper right plot shows the $m$ modes of the radial (conservative) component $F_r$, on a log-log scale. The dotted reference line is $\propto m^{-4}$. The lower plot shows the same data but rescaled by the expected power-law for a fourth-order puncture, i.e., $m^4 F_r^m$. We show a range of values of the Kerr parameter, $a/M = -0.9, -0.5, 0, 0.5, 0.9$, and we observe that rotation does not affect the convergence rate.}
\label{fig:mmode}
\end{figure}

To compute the total SF, we first compute the values for a range of $m$ modes (typically $m=0,\ldots,19$). As discussed in the Sec.~\ref{sec:intro}, for circular orbits the modes of the angular component $F_{\phi}$ (which is purely dissipative) are expected to converge exponentially fast, $F_{\phi}^m \sim \exp(-\lambda m)$. The upper left plot of Fig.~\ref{fig:mmode}, showing $-F_{\phi}^m$ as a function of $m$ on a semilog scale, confirms this expectation. 

By contrast, the modes of the radial component $F_{r}$ (which is purely conservative) are expected to exhibit power-law convergence, according to the rule $F_r^m \sim m^{-\zeta}$, where $\zeta = 2$ for 2nd- and 3rd-order punctures, and $\zeta=4$ for 4th-order punctures. In Paper I, we obtained results for punctures of 2nd, 3rd and 4th order, and demonstrated the expected convergence rates (see Figs.~13 and 14 in Paper I). In the Kerr case, we have only implemented the 4th-order puncture. The upper right plot of Fig.~\ref{fig:mmode} shows the magnitude of the modes, $|F_r^m|$, as a function of $m$, on a log-log scale, for $r_0 = 10M$ and a range of $a$. This graphically confirms that $F_r^m \sim m^{-4}$ in the large-$m$ limit, as expected. The bottom plot in Fig.~\ref{fig:mmode} shows the rescaled variable, $m^4 F_r^m$, which tends to a constant in the large-$m$ limit. We see that the value of the Kerr parameter $a$ does not affect the convergence rate.


\subsection{Self-force results}
Let us now present sample numerical results from our 4th-order puncture implementation.
In Tables \ref{tbl:Fr} and \ref{tbl:Fphi} we give the radial and angular components of the SF, $F_r^{\text{self}}$ and $F_\phi^{\text{self}}$, for circular orbits at three radii, $r_0 = 6M$, $10M$ and $\risco$.  Here $\risco$ denotes the radius of the innermost stable circular orbit, which is a root of the equation
\beq
r^2 - 6Mr - 3a^2 + 8 a \sqrt{Mr} = 0 .
\eeq
We give SF results for a range of Kerr parameters, $a/M = -0.9$, $-0.7$, $-0.5$, $0$, $0.5$, $0.7$ and $0.9$.

The lower value in each entry of Tables \ref{tbl:Fr} and \ref{tbl:Fphi} was obtained by Warburton and Barack \cite{Warburton:Barack:2010} using $l$-mode regularization in the frequency domain. The upper figure in each row was obtained via the $m$-mode scheme described in the previous sections, with figures in brackets indicating the estimated error bar on the last quoted decimal. (The results from \cite{Warburton:Barack:2010} are said to be 
accurate to all digits quoted.) Unsurprisingly (for our circular-orbit implementation), the time-domain figures are significantly less accurate than the frequency-domain results.

Each (upper) value in Tables \ref{tbl:Fr} and \ref{tbl:Fphi} was obtained by post-processing the results from multiple runs. First, we used runs of various resolutions to extrapolate to zero grid spacing (see Sec.~\ref{subsec:convergence}). For the modes $m=0$ (with the largest relaxation error) we further fitted the late-time data with a power-law relaxation model. To account for the large-$m$ tail (Sec.~\ref{subsec:modesums}), i.e., the modes $m>19$, we fitted the modes $m=12, \ldots, 19$ with a three-term model $Am^{-4} + Bm^{-5} + Cm^{-6}$. Estimates of the residual errors that remain after performing these steps were obtained. To estimate the total numerical error we combined the error estimates from discretization, relaxation, and tail-fitting steps in quadrature. 

\begin{table}
\begin{tabular}{| l | l c | l c | l c |}
\hline
\multicolumn{7}{|c|}{Radial component of SF, $(M^2/q^2) F_r^{\text{self}}$} \\
\hline
 & \multicolumn{2}{c|}{$r_0 = 6M$} & \multicolumn{2}{c|}{$r_0 = 10M$} & \multicolumn{2}{c|}{$r_0 = r_{\text{isco}}$} \\
\hline
\multirow{2}{*}{$a= -0.9M$} & \multicolumn{2}{c|}{--} & $\msp 4.941(1)$ & \multirow{2}{*}{$\e5$} & $\msp 9.6074(7)$ &  \multirow{2}{*}{$\e5$}  \\
	          &  \multicolumn{2}{c|}{--} & $\msp 4.939995$ &  & $\msp 9.607001$ &  \\
\hline
\multirow{2}{*}{$a= -0.7M$} & \multicolumn{2}{c|}{--} & $\msp 4.102(1)$ &  \multirow{2}{*}{$\e5$} & $\msp 1.1077(2)$ &  \multirow{2}{*}{$\e4$}  \\
	          &  \multicolumn{2}{c|}{--} & $\msp 4.100712$ &  & $\msp 1.107625$ &  \\
\hline
\multirow{2}{*}{$a= -0.5M$} & \multicolumn{2}{c|}{--} & $\msp 3.290(1)$ &  \multirow{2}{*}{$\e5$} & $\msp 1.2751(2)$ &  \multirow{2}{*}{$\e4$} \\
	          &  \multicolumn{2}{c|}{--} & $\msp 3.28942$ &  & $\msp 1.275170$ &  \\
\hline
\multirow{2}{*}{$a= \phantom{-} 0$} & $\msp 1.6771(2)$ & \multirow{2}{*}{$\e4$ } & $\msp 1.379(1)$ &  \multirow{2}{*}{$\e5$ } & $\msp 1.6771(2)$ & \multirow{2}{*}{$\e4$} \\
	          &  $\msp 1.677283$ & & $\msp 1.378448$ & & $\msp 1.677283$ &  \\
\hline
\multirow{2}{*}{$a= + 0.5M$} & $-2.423(4)$ & \multirow{2}{*}{$\e5$} & $-4.028(9)$ & \multirow{2}{*}{$\e6$} & $-6.925(5)$ & \multirow{2}{*}{$\e5$} \\
	          &  $-2.421685$ & & $-4.03517$ &  & $-6.922147$ &  \\
\hline
\multirow{2}{*}{$a= + 0.7M$} & $-9.530(3)$ & \multirow{2}{*}{$\e5$} & $-1.0913(9)$ & \multirow{2}{*}{$\e5$} & $-1.0886(4)$ & \multirow{2}{*}{$\e3$} \\
	          &  $-9.528095$ &  & $-1.091819$ &  & $-1.088457$ &  \\
\hline
\multirow{2}{*}{$a= + 0.9M$} & $-1.6458(5)$ & \multirow{2}{*}{$\e4$} & $-1.767(1)$ & \multirow{2}{*}{$\e5$} & $-1.1344(9)$ & \multirow{2}{*}{$\e2$}  \\
	          &  $-1.645525$ &  & $-1.768232$ & & $-1.133673$ &  \\
\hline
\end{tabular}
\caption{
Numerical results for the radial component $F_r^\text{self}$, for a range of Kerr rotation parameters $a$ and orbital radii $r_0$. Here $r_\text{isco}$ is the orbital radius of the innermost stable circular orbit (ISCO), which is a function of $a$ and takes the values $r_{\text{isco}} / M$ $= 8.717352, 8.142965, 7.554585, 6.0, 4.233003, 3.393128, 2.320883$ for $a / M = -0.9, -0.7, \ldots , 0.9$, respectively. The table compares the results of our implementation of $m$-mode regularization in the time domain using a 4th-order puncture (upper entries), with the results from the $l$-mode regularization method in the frequency domain from Warburton and Barack \cite{Warburton:Barack:2010, Warburton:Barack:2011} (lower entries). Parenthetical figures in the upper values indicate the estimated error bar on the last quoted decimals. All digits shown in the lower values are significant. Entries left empty correspond to orbits below the ISCO.}
\label{tbl:Fr}
\end{table}

\begin{table}
\begin{tabular}{| l | l c | l c | l c |}
\hline
\multicolumn{7}{|c|}{Angular component of SF, $-(M/q^2) F_\phi^{\text{self}}$} \\
\hline
 & \multicolumn{2}{c|}{$r_0 = 6M$} & \multicolumn{2}{c|}{$r_0 = 10M$} & \multicolumn{2}{c|}{$r_0 = r_{\text{isco}}$} \\
\hline
\multirow{2}{*}{$a= -0.9M$} & \multicolumn{2}{c|}{--} & $\msp 1.41470(1)$ & \multirow{2}{*}{$\e3$} & $\msp 2.18835(1)$ & \multirow{2}{*}{$\e3$} \\
	          &  \multicolumn{2}{c|}{--} & $\msp 1.414708$ &  & $\msp 2.188351$ &  \\
\hline
\multirow{2}{*}{$a= -0.7M$} & \multicolumn{2}{c|}{--} & $\msp 1.35624(1)$ & \multirow{2}{*}{$\e3$} & $\msp 2.57803(1)$ & \multirow{2}{*}{$\e3$} \\
	          &  \multicolumn{2}{c|}{--} & $\msp 1.356244$ &  & $\msp 2.578045$ &  \\
\hline
\multirow{2}{*}{$a= -0.5M$} & \multicolumn{2}{c|}{--} & $\msp 1.30226(1)$ & \multirow{2}{*}{$\e3$} & $\msp 3.08354(1)$ & \multirow{2}{*}{$\e3$} \\
	          &  \multicolumn{2}{c|}{--} & $\msp 1.302267$ &  & $\msp 3.083542$ &  \\
\hline
\multirow{2}{*}{$a= \msp 0$} & $\msp 5.304230(3)$ & \multirow{2}{*}{$\e3$} & $\msp 1.18592(1)$ & \multirow{2}{*}{$\e3$} & $\msp 5.30423(1)$ & \multirow{2}{*}{$\e3$}  \\
	          & $\msp 5.3042317$ &  & $\msp 1.185926$ &  & $\msp 5.304232$ &  \\
\hline
\multirow{2}{*}{$a= + 0.5M$} & $\msp 4.230745(3)$ & \multirow{2}{*}{$\e3$} & $\msp 1.09349(1)$ & \multirow{2}{*}{$\e3$} & $\msp 1.18357(4)$ & \multirow{2}{*}{$\e2$} \\
	          & $\msp 4.230749$ &  & $\msp 1.093493$ &  & $\msp 1.183567$ &  \\
\hline
\multirow{2}{*}{$a= + 0.7M$} & $\msp 3.928695(3)$ & \multirow{2}{*}{$\e3$} & $\msp 1.06216(1)$ & \multirow{2}{*}{$\e3$} & $\msp 1.94873(1)$ & \multirow{2}{*}{$\e2$} \\
	          & $\msp 3.928698$ &  & $\msp 1.062163$ &  & $\msp 1.948731$ &  \\
\hline
\multirow{2}{*}{$a= + 0.9M$} & $\msp 3.676723(8)$ & \multirow{2}{*}{$\e3$} & $\msp 1.03344(1)$ & \multirow{2}{*}{$\e3$} & $\msp 4.5079(2)$ & \multirow{2}{*}{$\e2$} \\
	          & $\msp 3.676726$ &  & $\msp 1.0334444$ &  & $\msp 4.508170$ &  \\
\hline
\end{tabular}
\caption{
Same as in Table \ref{tbl:Fr}, for the angular component $F_\phi^{\text{self}}$.
}
\label{tbl:Fphi}
\end{table}

We find that the results of the $m$-mode method are in good agreement with the frequency domain results of Ref.~\cite{Warburton:Barack:2010}, and the difference between the data sets lies broadly within the estimated error bars of our results. The error budget is dominated by discretization and relaxation errors, with tail-fitting error subdominant in our 4th-order puncture implementation. We hope that, in future, implementation of a truly 4th-order FD scheme (using 4th-order spatial differencing) will diminish the discretization error to the level where late-time relaxation error is left as the dominant error source. Methods for dealing with the latter (such as hyperboloidal slicing \cite{Zenginoglu:2011}, or multigrid refinement \cite{Dolan:Barack:2011}) are also ready for deployment. Hence it may be possible to arrive at a situation where tail-fitting is once again the dominant error source, in turn motivating the use of yet-higher-order punctures.

It is worth noting that, though $F_r^{\text{self}}$ is positive for all circular orbits on Schwarzschild (and all retrograde orbits on Kerr), it is negative for some prograde orbits on Kerr. This phenomenon was clearly illustrated in Fig.~5 of \cite{Warburton:Barack:2010}, and is evident in the results of Table~\ref{tbl:Fr}. The condition $F_r^{\text{self}} = 0$ describes a curve in the $r_0$--$a$ plane. We note that the relative accuracy of the $m$-mode sum result diminishes for $r_0$ and $a$ near this curve, as the individual $m$ mode contributions conspire to cancel (though the sum may be zero, the individual $m$-modes are not). This means that small relative errors in individual $m$ modes become magnified into large relative errors in the mode sum. In Paper I, we described this numerical problem, so-called ``mode cancellation error'', in the context of orbits at large radii.



\section{Outlook}\label{outlook}

This paper reports on a first time-domain calculation of the self force in Kerr spacetime. Our purpose here was to develop a computational infrastructure (puncture functions, numerical methods and a working code) suitable for the Kerr problem, and illustrate the efficacy of our $m$-mode regularization approach. 

There are a number of obvious ways in which the efficiency of our numerical method may be improved. One may implement a higher-order-convergent FD scheme, and/or use a higher order ($n>4$) puncture. The former improvement would lead to more rapid convergence with respect to the FD grid spacing, while the latter would accelerate the convergence of the $m$-mode sum. We recall our result from Paper I, according to which the individual modal contributions to the SF fall off at large $m$ as $m^{-n}$ for even order $n$, but only as $m^{-n+1}$ for odd order $n$. Hence, it would be desirable to go directly to $n=6$. The derivation of such a high-order puncture is possible by extending the formulation of Sec.\ \ref{subsec:punc} to higher order using the methods developed in Ref.\ \cite{Ottewill:Wardell}; such an extension would be straightforward in principle, even if tedious in practice. Moving to a higher-order-convergent FD scheme would necessitate either (i) dealing more carefully with the nonsmoothness of the residual field along the worldline, or (ii) using a higher-order puncture in order to improve the differentiability of this field.  Other possible ways to improve the efficiency of the numerical method include (i) implementing a mesh-refinement algorithm such as the one described in Paper I (or the more systematic algorithm developed by Thornburg \cite{Thornburg:2009, Thornburg:2010}), or (ii) improving the treatment of boundary conditions, e.g., via the hyperbolic slicing method of Ref.\  \cite{Zenginoglu:2011}.

Several natural extensions of this work are possible. First, one may consider more generic orbits: eccentric, inclined, or even unbound; our time-domain framework has the advantage that it requires only a little adaptation to accommodate different classes of orbits. The puncture formulation described in Sec.\ \ref{subsec:punc} is applicable for any type of motion in Kerr spacetime, with the specific orbital details entering only through the explicit values of the coefficients of $\esq_{(n)}$, $\alpha_{(n)}$ and $\beta_{(n)}$. The latter are easily determined for any given orbit based on the covariant expressions (\ref{epsilon}), (\ref{alpha}) and (\ref{beta}). The generalization of our FD algorithm to cope with generic orbits is also rather straightforward: With a 4th-order puncture implementation, the residual field is twice differentiable at the particle (cf.\ Table I of Paper I), and---better still---its individual $m$-modes, which are the numerical variables, are {\em thrice} differentiable there (and smooth anywhere else). This means that grid points in the vicinity of the worldline would not require any special treatment within the FD scheme even when the orbits are non-circular (as long as one is content with a 2nd-order-convergent FD scheme, as we do in the current work). This stands in sharp contrast to the situation in the 1+1D framework of Ref.\ \cite{Barack:Sago:2010}, where the lack of differentiability of the numerical variables made the FD scheme significantly more complicated in the case of eccentric orbits, and the generalization from circular orbits required much further development.

Once a generic-orbit scalar-field SF code is at hand, one may attempt another important extension: the inclusion of the back-reaction effect of the SF on the orbital motion, to compute the evolving orbit in a self-consistent manner. The time-domain framework allows, in principle, to correct the motion ``in real time'' as the evolution proceeds. However, it is a non-trivial task to formulate an algorithm for incorporating the SF information in a manner that is both physically robust and computationally efficient. The scalar-field SF problem offers a convenient environment for test and development, as it is devoid of gauge-related complexities. A self-consistent evolution has not been attempted so far, neither for the gravitational SF nor for the scalar-field SF.

Of course, the most important extension of our code is to the gravitational case. We have already begun work to apply the strategy of $m$-mode regularization to the gravitational SF. The generalization of the puncture formulation to the gravitational case is very straightforward, and gravitational-field punctures akin to $\Phi_{\cal P}^{[n]}$ are already available through 4th-order. Our numerical strategy, following \cite{Barack:Sago:2010,Barack:Sago:2007}, is to work directly with the components of the Lorenz-gauge metric perturbation as numerical variables. The linearized Einstein equations are decomposed into $m$-modes, yielding (for each $m$) a set of 10 coupled partial differential equations in 2+1D for the various coordinate components of the $m$-decomposed metric. The key element of our method involves a judicious use of the ($m$-decomposed) Lorenz gauge conditions in order to reformulate the set of 2+1D perturbation equations so that they are amenable to numerical time-evolution and so that gauge-condition violations are automatically suppressed during the evolution. Care should also be taken to ensure that suitable boundary conditions are used for the various components at the poles. However, the numerical algorithm is otherwise similar to the one developed in the current work. We have already been able to successfully implement this approach in the test case of circular orbits on Schwarzschild, reproducing some of the gravitational SF results of Ref.\ \cite{Barack:Sago:2007}. 

However, using our method, we have thus far only been able to solve for the modes $m\geq 2$. The two lowest modes, $m=0$ and $m=1$, which contain nonradiative monopole and dipole pieces, do not evolve stably using our numerical scheme: for these two modes we find our numerical solutions to be growing linearly in time (even though gauge violations remain small). This situation is similar to the one described in the 1+1D framework of Refs.\ \cite{Barack:Sago:2010,Barack:Sago:2007}, where no stable evolution scheme has been found for the monopole ($l=0$) and dipole ($l=1$) modes. Instead, Refs.\ \cite{Barack:Sago:2010,Barack:Sago:2007} had to rely on frequency-domain methods to obtain the monopole and dipole contributions. Our ambition is to find a method for calculating the two problematic modes ($m=0$, $1$, in our case) using time-evolution in 2+1D, relieving us from the need to resort to frequency-domain methods, the latter being useful only for bound orbits with sufficiently small eccentricity. We have made significant progress towards this end, and will report our findings in a forthcoming third paper of this series, to be focused on the gravitational SF problem in Kerr.

\acknowledgments

SD acknowledges support from EPSRC through Grant No.~EP/G049092/1. LB acknowledges support from STFC through Grant No.~PP/E001025/1. We are grateful for the use of the {\sc Iridis} 3 HPC facility at the University of Southampton.

\appendix

\section{Coefficients for puncture functions through fourth order}
\label{appendix:4thord}
Here we give explicit expressions for the coefficients ${\sco}_{ijk}$ appearing in Eq.~(\ref{esqn}), specialized to circular equatorial geodesic orbits of Boyer-Lindquist radius $r_0$ about a Kerr black hole of mass $M$ and spin parameter $a$. We give all coefficients necessary for computing $\sco_{(n)}$ through $n=5$. We also give explicit expressions for the functions $\alpha_{(n)}$ and $\beta_{(n)}$ through $n=5$. This constitutes sufficient input for constructing the 2nd, 3rd and 4th-order punctures through Eqs.\ (\ref{eq:punc-2nd})--(\ref{eq:punc-4th}).
For compactness, we give the coefficients in terms of the dimensionless quantities 
\beq
\tilde r_0 \equiv r_0/M, \quad\quad
\tilde a \equiv a/M, \quad\quad
\eeq
in terms of which we have 
\beq
v=\tilde r_0^{-1/2}, \quad\quad
\tilde{\Delta}_0 \equiv M^{-2}\Delta(r_0) = \tilde r_0^2-2\tilde r_0+\tilde a^2,
\eeq
as well as
\begin{equation}
\tilde{u}^\phi \equiv M u^\phi = \frac{1}{\tilde r_0\sqrt{\tilde{r}_0-3+2\tilde{a}v }},
\quad \quad
u^t = \frac{v(\tilde r_0+\tilde{a}v)}{\sqrt{\tilde{r}_0-3+2\tilde{a}v }}.
\end{equation}

Let us begin with the coefficients ${\sco}_{ijk}$ of the $\epsilon^2$-expansion (\ref{esqn}). The non-zero coefficients at $O(\delta x^2)$ are given by  
\begin{gather}
{\sco}_{002}=\tilde{\Delta }_0 (\tilde{u}^{t})^2, \quad\quad
{\sco}_{020}=\tilde{r}_0 {}^2, \quad\quad
{\sco}_{200}=\frac{\tilde{r}_0{}^2}{\tilde{\Delta }_0},
\end{gather}
at $O(\delta x^3)$ by
\begin{gather}
{\sco}_{102}=(\tilde{r}_0-1) (u^t)^2,\quad\quad
{\sco}_{120}=\tilde{r}_0, \quad\quad
{\sco}_{300}=\frac{\tilde{r}_0 (\tilde{a}^2-\tilde{r}_0)}{\tilde{\Delta }_0{}^2},
\end{gather}
at $O(\delta x^4)$ by
\begin{gather}
{\sco}_{220}=-\frac{\tilde{r}_0-3 \tilde{a}^2}{6 \tilde{\Delta }_0},\quad\quad
{\sco}_{040}=\frac{1}{12} \Big[3 \tilde{a}^2-\tilde{r}_0 (\tilde{r}_0-2) \Big],\quad \quad
{\sco}_{400}=\frac{2 \tilde{a}^2 \tilde{r}_0 (1-6 \tilde{r}_0) +3 \tilde{a}^4+\tilde{r}_0{}^2 (8 \tilde{r}_0-1)}{12 \tilde{\Delta }_0{}^3}, \nonumber \\
{\sco}_{004}=\frac{\tilde{\Delta }_0 (u^t)^2}{12\tilde{r}_0{}^4}  \big(\tilde{a}v -\tilde{r}_0 \big) \Big[\tilde{a} (\tilde{r}_0-5) v^{-1}+2 \tilde{a}^2+\tilde{r}_0{}^2 (\tilde{r}_0+1)\Big] ,\nonumber \\
{\sco}_{022}=-\frac{(\tilde{u}^{\phi })^2}{6 \tilde{r}_0}\Big[\tilde{a}^4 (9 \tilde{r}_0+11)+2 \tilde{a}^3 v^{-1} (\tilde{r}_0+1) (3 \tilde{r}_0-8)+\tilde{a}^2 \tilde{r}_0 (3\tilde{r}_0{}^3 + 10\tilde{r}_0{}^2 - 3\tilde{r}_0+2)\nonumber \\
+2 \tilde{a} v^{-5} (3 \tilde{r}_0{}^2-11 \tilde{r}_0+2)+\tilde{r}_0{}^4 (\tilde{r}_0-1) (3 \tilde{r}_0-2)\Big] ,\nonumber \\
{\sco}_{202}=\frac{(\tilde{u}^{\phi })^4 }{6 v^2 \tilde{\Delta }_0} \left(\tilde{r}_0-3+2\tilde a v\right) \Big[6 \tilde{\Delta }_0 (\tilde{a}+v^{-3}){}^2+\tilde{a}^4 (9 \tilde{r}_0+7)+2 \tilde{a}^3 v^{-1} (3 \tilde{r}_0 {}^2-7\tilde{r}_0 -4) \nonumber \\
+\tilde{a}^2 \tilde{r}_0 (\tilde{r}_0 {}^2-1) (3 \tilde{r}_0+5)-14 \tilde{a} v^{-5}(\tilde{r}_0-1) -\tilde{r}_0{}^4(\tilde{r}_0-1) \Big],
\end{gather}
and, finally, at $O(\delta x^5)$, they are given by
\begin{gather}
{\sco}_{140}=\frac{1}{12} (1-\tilde{r}_0), \quad\quad
{\sco}_{320}=\frac{\tilde{a}^2 (5-6 \tilde{r}_0)+\tilde{r}_0{}^2}{12 \tilde{\Delta }_0{}^2}, \nonumber\\
{\sco}_{500}=\frac{\tilde{a}^4 (4-9 \tilde{r}_0)+\tilde{a}^2 \tilde{r}_0 (12 \tilde{r}_0{}^2-5\tilde{r}_0+3)-\tilde{r}_0{}^2 (6 \tilde{r}_0{}^2-2 \tilde{r}_0+1)}{12 \tilde{\Delta }_0{}^4}, \nonumber \\
{\sco}_{122}=\frac{(\tilde{u}^{\phi })^2}{12 \tilde{r}_0{}^2}\Big[\tilde{a}^4 (9 \tilde{r}_0+29)-2 \tilde{a}^3 v^{-1}(5 \tilde{r}_0+16)+\tilde{a}^2 \tilde{r}_0 (3\tilde{r}_0-17 \tilde{r}_0{}^2+12) \nonumber \\
+2 \tilde{a} \tilde{r}_0{}^3 v^{-1}(11-6 \tilde{r}_0) -2 \tilde{r}_0{}^4 (3\tilde{r}_0{}^2+2 \tilde{r}_0-3)\Big], \nonumber \\
{\sco}_{104}=-\frac{(\tilde{u}^{\phi })^2}{12\tilde{r}_0{}^4}\left(1+\tilde{a}v^3\right)\Big[\tilde{r}_0{}^6 (\tilde{r}_0{}^2 +4\tilde{r}_0-9)+\tilde{a} v^{-9} (\tilde{r}_0{}^2-8\tilde{r}_0+9)+\tilde{a}^2 \tilde{r}_0{}^3 (8 \tilde{r}_0{}^2 +3\tilde{r}_0-21) \nonumber \\
-\tilde{a}^3 v^{-3} (\tilde{r}_0+3) (6 \tilde{r}_0-11)+\tilde{a}^4 \tilde{r}_0 (3 \tilde{r}_0{}^2 +15 \tilde{r}_0-10)
-\tilde{a}^5 v^{-1} (3 \tilde{r}_0+19)+6 \tilde{a}^6\Big] , \nonumber \\
{\sco}_{302}=\frac{(\tilde{u}^{\phi })^4}{12 \tilde{\Delta }_0{}^2}\left(\tilde{r}_0-3+2\tilde a v\right) \Big[\tilde{r}_0{}^6(\tilde{r}_0-1)-4 \tilde{\Delta }_0 \tilde{r}_0{}^4+\tilde{a}^2 \tilde{r}_0 [4 \tilde{\Delta }_0 ( 3 \tilde{r}_0{}^2-4\tilde{r}_0+5)- 6 \tilde{r}_0{}^5 \nonumber \\
-7\tilde{r}_0{}^4 +19\tilde{r}_0{}^3 +6\tilde{r}_0{}^2 -6\tilde{r}_0]-\tilde{a}^6 (3 \tilde{r}_0+7)+2 \tilde{a}^5 v^{-1} (\tilde{r}_0+4)-\tilde{a}^4 (23 \tilde{r}_0{}^3-\tilde{r}_0{}^2-26\tilde{r}_0 +12 \tilde{\Delta }_0) \nonumber \\
 +4 \tilde{a}^3 v^{-1} (2 \tilde{\Delta }_0-3 \tilde{r}_0{}^4+10 \tilde{r}_0{}^3-8 \tilde{r}_0)+2 \tilde{a} v^{-5} [8 \tilde{\Delta }_0+\tilde{r}_0 (13 \tilde{r}_0{}^2-22\tilde{r}_0 +6-12 \tilde{\Delta }_0)]\Big].
\end{gather}

For the 3rd-order puncture [Eq.\ (\ref{eq:punc-3rd})] we additionally require the function $\alpha_{(4)}$, and for the 4th-order puncture [Eq.\ (\ref{eq:punc-4th})] we require $\alpha_{(5)}$ and $\beta_{(5)}$. These are compactly given in the form 
\begin{align}
\alpha_{(4)} =& -\frac{M^2 (\tilde{u}^{\phi })^2 }{6 \tilde{\Delta }_0{}^4 \tilde{r}_0{}^6}
   \Big\{\tilde{\Delta }_0 \tilde{r}_0{}^4 \delta \tilde{r}^2+\tilde{\Delta }_0{}^2 \tilde{r}_0{}^4 \delta \theta ^2 
   + \tilde{\Delta }_0{}^2 \tilde{r}_0 [\tilde{a}^2 (\tilde{r}_0+2)+\tilde{r}_0{}^3]\delta \phi ^2 \Big\}
   \nonumber \\ & \times
   \Big[\tilde{\Delta }_0 \tilde{r}_0{}^4
   (2 \tilde{a} v^{-1}-3 \tilde{a}^2+3\tilde{r}_0-2 \tilde{r}_0{}^2 )\delta \tilde{r}^2
   + \tilde{\Delta }_0{}^2 \tilde{r}_0{}^4 (-4
   \tilde{a} v^{-1}+3 \tilde{a}^2+\tilde{r}_0{}^2)\delta \theta ^2
   \nonumber \\ &
   +\tilde{\Delta }_0{}^3 \tilde{r}_0 
   (\tilde{a}+v^{-3}){}^2 \delta \phi ^2 
   \Big],
\end{align}
\begin{align}\label{alpha5}
\alpha_{(5)} =& -\frac{M^2 (\tilde{u}^{\phi })^2 }{6 \tilde{\Delta }_0{}^4 \tilde{r}_0{}^6}
   \Big\{\tilde{\Delta }_0 \tilde{r}_0{}^4 \delta \tilde{r}^2+\tilde{\Delta }_0{}^2 \tilde{r}_0{}^4 \delta \theta ^2 
   + \tilde{\Delta }_0{}^2 \tilde{r}_0 [\tilde{a}^2 (\tilde{r}_0+2)+\tilde{r}_0{}^3]\delta \phi ^2
   + [ \tilde{r}_0{}^3 (\tilde{a}^2-\tilde{r}_0)\delta \tilde{r}^2
   \nonumber \\ &
   +\tilde{\Delta }_0{}^2 \tilde{r}_0{}^3 \delta \theta ^2 + \tilde{\Delta }_0{}^2 (\tilde{r}_0{}^3
   -\tilde{a}^2)\delta \phi ^2 ] \delta \tilde{r}
   \Big\}\Big\{\tilde{\Delta }_0 \tilde{r}_0{}^4
   (2 \tilde{a} v^{-1}-3 \tilde{a}^2+3\tilde{r}_0-2 \tilde{r}_0{}^2 )\delta \tilde{r}^2
   \nonumber \\ &
   + \tilde{\Delta }_0{}^2 \tilde{r}_0{}^4 (-4
   \tilde{a} v^{-1}+3 \tilde{a}^2+\tilde{r}_0{}^2)\delta \theta ^2
   +\tilde{\Delta }_0{}^3 \tilde{r}_0 
   (\tilde{a}+v^{-3}){}^2 \delta \phi ^2 
   + \Big[\tilde{\Delta }_0{}^2 \tilde{r}_0{}^3 (9
   \tilde{a}^2-10 \tilde{a} v^{-1}\nonumber \\ &
   + 4 \tilde{r}_0{}^2-3\tilde{r}_0)\delta \theta ^2 + \tilde{\Delta }_0{}^2 \Big(\tilde{a}^2
   \tilde{r}_0 (3 \tilde{r}_0{}^2-2 \tilde{r}_0+5)+2 \tilde{a} v^{-5}(\tilde{r}_0-1) -3
   \tilde{a}^4+\tilde{r}_0{}^4 (4 \tilde{r}_0-7)\Big)\delta \phi ^2
   \nonumber \\ &
   + \tilde{r}_0{}^3 (2 \tilde{a}^3
   v^{-1}-2 \tilde{a}^2 \tilde{r}_0(\tilde{r}_0-3) -2 \tilde{a} v^{-3}-3
   \tilde{a}^4+2\tilde{r}_0{}^3 -3\tilde{r}_0{}^2)\delta \tilde{r}^2\Big]\delta \tilde{r}\Big\},
\end{align}
and
\begin{align}
\beta_{(5)} =&\frac{ M^2 \delta \tilde{r}}{8 \tilde{\Delta }_0{}^2 \tilde{r}_0{}^7} \Big\{
   \tilde{r}_0{}^2\Big[ \tilde{r}_0{}^3 \delta \tilde{r}^2 + \tilde{\Delta }_0 \tilde{r}_0{}^3 \delta \theta ^2 +\tilde{\Delta }_0 \Big(\tilde{a}^2 (\tilde{r}_0+2)+\tilde{r}_0{}^3\Big)\delta \phi ^2\Big] \Big[ \Big(
   8 \tilde{a} \tilde{r}_0 \tilde{u}^{\phi }u^t(\tilde{r}_0-1) +10  \tilde{a}^3 u^t \tilde{u}^{\phi } 
   \nonumber \\ &
   + \tilde{a}^2 [2 \tilde{r}_0 (\tilde{u}^{\phi })^2(2-3 \tilde{r}_0) -5 (u^t)^2]
   -5 \tilde{a}^4 (\tilde{u}^{\phi })^2-\tilde{r}_0 [\tilde{r}_0{}^3 (\tilde{u}^{\phi })^2+2 (u^t)^2 (\tilde{r}_0-2)]\Big)\delta \tilde{r}^2  
   \nonumber \\ &
   +\tilde{\Delta }_0 \Big(15 \tilde{a}^4 (\tilde{u}^{\phi })^2 -30 \tilde{a}^3 u^t \tilde{u}^{\phi }+\tilde{a}^2 [\tilde{r}_0 (19 \tilde{r}_0-6) (\tilde{u}^{\phi })^2+15 (u^t)^2]+2  \tilde{a} \tilde{r}_0 \tilde{u}^{\phi }u^t (6-11 \tilde{r}_0)
   \nonumber \\ &
   +\tilde{r}_0 [4 \tilde{r}_0{}^3 (\tilde{u}^{\phi })^2+3 (u^t)^2 (\tilde{r}_0-2)]\Big)\delta \theta ^2 +\tilde{\Delta }_0{}^2 u^t \Big(3 u^t-2 \tilde{a} \tilde{u}^{\phi }\Big)\delta \phi ^2\Big] - 2 \tilde{\Delta }_0{}^2 u^t \tilde{u}^{\phi } \Big[\tilde{a}^2 \tilde{u}^{\phi } (\tilde{r}_0+2) 
   \nonumber \\ &
   -2 u^t \tilde{a}+\tilde{r}_0{}^3 \tilde{u}^{\phi }\Big] \Big[u^t-\tilde{a} \tilde{u}^{\phi }\Big] \Big[3 \tilde{r}_0{}^4 \delta \tilde{r}^2 +3 \tilde{\Delta }_0 \tilde{r}_0{}^4 \delta \theta ^2 +\tilde{\Delta }_0 \Big(\tilde{a}^2 [4 \tilde{r}_0{}^3 (\tilde{u}^{\phi })^2(\tilde{r}_0+2)+3 \tilde{r}_0{}^2
   +6 \tilde{r}_0
   \nonumber \\ &
   +8 (u^t)^2]-8 u^t \tilde{a}^3 (\tilde{r}_0+2) \tilde{u}^{\phi }-8 u^t \tilde{a} \tilde{r}_0{}^3 \tilde{u}^{\phi } +2 \tilde{a}^4 (\tilde{r}_0+2){}^2 (\tilde{u}^{\phi })^2
   +\tilde{r}_0{}^4 [2 \tilde{r}_0{}^2 (\tilde{u}^{\phi })^2+3]\Big)\delta \phi ^2 \Big]\delta \phi ^2\Big\}.
\end{align}
We note that our expression for $\alpha_{(5)}$ contains terms which are of $O(\delta x^6)$ and therefore it is not strictly consistent with our original definition [a Taylor series truncated at $O(\delta x^5)$]. However, the inclusion of these terms allows us to express $\alpha_{(5)}$ more compactly, and we opt to retain them despite the slight abuse of notation. Of course, the additional $O(\delta x^6)$ terms do not affect the 4th-order nature of the puncture. The 4th-order puncture employed in our numerical implementation uses $\alpha_{(5)}$ as it is given in Eq.\ (\ref{alpha5}). A {\sc Mathematica} workbook with details of the fourth-order puncture function is available \cite{Wardell-mathematica}.

\section{Second-order effective source}
\label{appendix:2nd-ord}
For the 2nd-order effective source, defined by $\Seff^{[2]} = S(x) - \Box \Phipunc^{[2]}$, we obtain the $m$-mode decomposition
\beq
\Seff^{[2]m}(t,r,\theta) = \frac{q}{2\pi} e^{-im (\Omega t + \Delta \phi)} \sum_{k=1}^5 S_k I_k^m, 
\label{Seff-n2}
\eeq
where $\Delta \phi$ was defined in Eq.~(\ref{delphi-def}) and
the integrals $I_1^m, \ldots, I_5^m$ are defined and expressed in terms of elliptic integrals in Eqs.~(B9)--(B13) of Paper I. Since there is the possibility of notional confusion, let us give one of these integrals explicitly:
\begin{eqnarray}
I_1^m &\equiv& \int_{-\pi}^\pi \esq_{(3)}^{-3/2} e^{-i m \delta \phi} d (\delta \phi) \nonumber \\
 &=& \frac{\gam}{M^3 B^{3/2}} \left[ p_{1K}^m(\rhohat) K(\gam) + \rhohat^{-2} p_{1E}^m(\rhohat) E(\gam) \right],
\end{eqnarray}
where $\esq_{(3)}$ is defined in Eq.~(\ref{esqn}), $B$ is given in Eq.\ (\ref{AB}), and $\rhohat$ and $\gamma$ are given in Eq.~(\ref{rhohat-gamma-def}). The polynomials $p_{1K}^m$, $p_{1E}^m$ for $m=0,\ldots,5$ are given in Table II of Ref.~\cite{Barack:Golbourn:2007}; higher-order terms can be computed easily with a symbolic algebra package.

The quantities $S_k$ in (\ref{Seff-n2}) are Kerr-generalizations of the ones given in Paper I. They read 
\begin{eqnarray}
 S_1 &=& M^2 \rho^{-2} [ (\tilde r-1)X +\tilde{\Delta} ({\sco}_{200} + 3{\sco}_{300} \delta \tilde{r}) + (1 + \dth \cot \theta ) ({\sco}_{020} + {\sco}_{120} \delta \tilde{r} ) ],\\
 S_2 &=& M^2 \rho^{-2} \Big[ \left( \gamma^{\phi\phi} - 2\tilde{\Omega} \gamma^{t\phi} + \tilde{\Omega}^2 \gamma^{tt}  \right) B -  2(\tilde r-1) {\sco}_{102} \Big], \\
 S_3 &=& -3 M^4 \rho^{-2} \Big[\tilde{\Delta} \left( \frac{X}{2} - {\sco}_{102} \right)^2 +\Big({\sco}_{020} + {\sco}_{120} \delta \tilde{r} \Big)^2 \dth^2 \Big], \\
 S_4 &=& 3 M^4 \rho^{-2} \Big[\tilde{\Delta} {\sco}_{102}^2- \left( \gamma^{\phi\phi} - 2\tilde\Omega \gamma^{t\phi} + \tilde\Omega^2 \gamma^{tt}  \right) B^2 \Big], \\
 S_5 &=& -3 M^4 \rho^{-2} \Big[\tilde{\Delta} \left( \frac{X}{2} + {\sco}_{102} \right)^2 +\Big({\sco}_{020} + {\sco}_{120} \delta \tilde{r} \Big)^2 \dth^2 \Big],
\end{eqnarray}
where the coefficients ${\sco}_{ijk}$ are the ones given in Appendix \ref{appendix:4thord}, a tilde denotes a-dimensionalization using $M$ (e.g., $\tilde r=r/M$ or $\tilde\Omega=M\Omega$), $\rho^2$ is defined in (\ref{rho-Delta-def}), and we have introduced
\begin{equation}
X = 2 {\sco}_{200} \delta \tilde{r} + 3 {\sco}_{300} \delta \tilde{r}^2 + {\sco}_{120} \dth^2 + 2{\sco}_{102},
\end{equation}
\begin{eqnarray}
\gamma^{tt} &=& -\left[ \frac{(\tilde{r}^2+\tilde{a}^2)^2}{\tilde{\Delta}} - \tilde{a}^2 \sin^2 \theta \right] ,  \\ 
\gamma^{t\phi} &=& -\frac{2 \tilde{a} \tilde{r}}{\tilde{\Delta}},  \\ 
\gamma^{\phi \phi} &=& \frac{1}{\sin^2 \theta} - \frac{\tilde{a}^2}{\tilde{\Delta}} .
\end{eqnarray}
Finally, $S_{\Psi}^m$ featuring in Eq.~(\ref{Boxm}) is given by $- r \Delta \rho^2 \Sigma^{-2} \Seff^{[2]m}$. 

\end{document}